\documentstyle[11pt,aaspp4,flushrt]{article}  
\received{RECEIPT DATE 1996}
\revised{REVISION DATE 1996}
\accepted{ACCEPT DATE 1996}
\journalid{VOL}{JOURNAL DATE 1996}
\articleid{START PAGE}{END PAGE}
          \font\sixrm=cmr6       
                     
\def\pr{Phys. Rev.}                             % DO NOT DELETE
\def\app{Astroparticle Phys.}                   % DO NOT DELETE
\def\asr{Adv. Space Res.}                       % DO NOT DELETE
\def\eos{EOS}                                   % DO NOT DELETE
\def\grl{Geophys. Res. Lett.}                   % DO NOT DELETE
\def\jgr{J. Geophys. Res.}                      % DO NOT DELETE
\def\pss{Planet. Sp. Sci.}                      % DO NOT DELETE
\def\rmp{Rev. Mod. Phys.}                       % DO NOT DELETE
\def\sp{Solar Phys.}                            % DO NOT DELETE
\def\ssr{Space Sci. Rev.}                       % DO NOT DELETE
\def\lefthead{Cosmic Rays from Gas and Dust}
\def\righthead{Ellison, Drury, and Meyer}
\def\QG      {Q_{\rm G}}
\def\AG      {A_{\rm G}}
\def\MG      {M_{\rm G}}
\def\lossmom {t_{\rm loss,mom}}
\def\vGmax   {v_{\rm G,max}}
\def\lappeq{\lower 0.5ex\hbox{$\; \buildrel < \over \sim \;$} 
               \allowbreak}
\def\gappeq{\lower 0.5ex\hbox{$\; \buildrel > \over \sim \;$} 
               \allowbreak}

\def\ct      {$^{12}$C\ }

\def\os      {$^{16}$O\ }

\def\nett    {$^{22}$Ne\ }

\def\cto     {$^{12}$C}

\def\oso     {$^{16}$O}

\def\netto   {$^{22}$Ne}

\def\Tc   {$T_{\rm c}$\ }
\def\Tco  {$T_{\rm c}$}
\def\lesssim{\mathrel{\hbox{\rlap
  {\hbox{\lower4pt\hbox{$\sim$}}}\hbox{$<$}}}}
\def\gtrsim{\mathrel{\hbox{\rlap
  {\hbox{\lower4pt\hbox{$\sim$}}}\hbox{$>$}}}}
\def\eqb{\begin{equation}}
\def\eqe{\end{equation}}
\def\cosray{cosmic ray }
\def\crs{cosmic rays }
\def\Vsk{V_{\rm sk}}
\def\Ux{U_{\rm x}(x)}
\def\Rsk{R_{\rm sk}}
\def\x#1{\times 10^{#1}}
\def\Z{\!\!}
\def \kmps{km s$^{-1}$}
\def\pcc{cm$^{-3}$}
\def\Hetwo{He$^{+2}$ }
\def\Tbn{\Theta_{\rm Bn}}
\def\dFEB{d_{\rm FEB}}
\def\vG{v_{\rm G}}

\def\acctime{\tau_{\rm a}}
\def\accscale{t_{\rm acc}}
\def\tSNR{t_{\rm SNR}}
\def\TG{T_{\rm G}}

\def\losssput{t_{\rm loss,sput}} 
\def\qsput{q_{\rm sput}} 
\def\betaG{\beta_{\rm G}}
\def\betaGmax{\beta_{\rm G,max}}
\def\denH{n_{\rm H}}
\def\denHe{n_{\rm He}}
\def\denFe{n_{\rm Fe}}
\def\denGr{n_{\rm G}}
\def\denGrFe{n_{\rm G,Fe}}
\def\dengas{\denH}

\def\mp{m_{\rm p}}
\def\EnSN{E_{\rm SN}}
\def\rg{r_{\rm g}}

\def\Tbn{\Theta_{\rm Bn}}

\def\dFEB{d_{\hbox{\sixrm FEB}}}

\def\L{\lambda}
\def\Lz{\lambda_0}
\def\rg{r_{\rm g}}
\def\rgone{r_{\rm g1}}
\def\Emax{E_{\rm max}}
\def\Epmax{E_{\rm p,max}}
\def\mc{Monte Carlo }
\def\ie{i.e. }
\def\iec{i.e., }
\def\eg{e.g. }
\def\egc{e.g., }
\def\etal{et al. }
\def\Phialpobs{\Phi_{\rm \alpha,obs}}
\def\Phialps{\Phi_{\rm \alpha,s}}
\def\Phipobs{\Phi_{\rm p,obs}}
\def\Phips{\Phi_{\rm p,s}}

\def\Aa{A_{\!\alpha}}
\def\Qa{Q_{\!\alpha}}
\def\ALS{Axford \etal  }

\def\BO{Blandford and Ostriker }
\def\DAV{Drury, Aharonian, and V\"olk }
\def\EBJ{Ellison, Baring, and Jones }
\def\EE{Ellison and Eichler }
\def\EGBS{Ellison \etal } %Ellison, Giacalone, Burgess, and Schwartz

\def\EJE{Ellison, Jones, and Eichler }
\def\EJR{Ellison, Jones, and Reynolds }

\def\EMP{Ellison, M\"obius, and Paschmann }

\def\JE{Jones and Ellison }

\begin{document}
\title{Galactic Cosmic rays from Supernova Remnants:
II Shock  Acceleration of Gas and Dust}
   \author{Donald C. Ellison}                      %  AASTEX format only
   \affil{Department of Physics, North Carolina State University, \\
      Box 8202, Raleigh NC 27695, U.S.A.\\
      don\_ellison@ncsu.edu}
   \author{Luke O'C. Drury}
   \affil{Dublin Institute for Advanced Studies, 
School of Cosmic Physics, \\
5 Merrion Square, Dublin 2, IRELAND \\ 
ld@cp.dias.ie}
\author{Jean-Paul Meyer}
\affil{Service d'Astrophysique, CEA/DSM/DAPNIA \\
Centre d'Etudes de Saclay, 91191 Gif-sur-Yvette, FRANCE \\
meyer@sapvxb.saclay.cea.fr}
   \authoraddr{Department of Physics, North Carolina State University,
       Box 8202, Raleigh NC 27695, U.S.A.}
\vskip18pt
\centerline{Submitted to the {\it Astrophysical Journal,} October 8,
1996}
\vskip6pt
\centerline{Revised, January 20, 1997}
\vskip6pt
\centerline{Accepted, January 24, 1997}
\begin{abstract}
We present a quantitative model of galactic cosmic ray (GCR) 
origin and acceleration, wherein a mixture of interstellar  and/or 
circumstellar gas and dust is accelerated by a supernova remnant 
(SNR) blast wave.
The gas and dust are accelerated simultaneously, but differences in 
how each component is treated by the shock leaves a distinctive 
signature  which we believe exists in the cosmic ray composition data.
A re-examination of the detailed GCR elemental composition, 
presented in a companion paper, has led us to abandon the long held 
assumption that GCR abundances are somehow determined by 
first ionization potential (FIP).  
Instead, volatility and mass (presumably mass-to-charge ratio) seem to
better organize the data: among the volatile elements, the abundance
enhancements relative to solar increase with mass (except for the
slightly high H/He ratio); the more refractory elements seem
systematically overabundant relative to the more volatile ones in a
quasi-mass-independent fashion.
If this is the case, material locked in grains in the interstellar 
 medium must be accelerated to cosmic ray energies more 
efficiently than interstellar gas-phase ions.
Here we present results from a nonlinear shock model which includes 
(i) the direct acceleration of interstellar  gas-phase ions, 
(ii) a simplified model for the direct acceleration of  weakly 
charged grains  to $\sim 100$~keV/amu energies, simultaneously 
with the acceleration of  the gas ions,
(iii)  the energy losses  of grains  colliding with 
the ambient gas,
(iv) the sputtering of grains, and 
(v) the simultaneous acceleration of the sputtered ions to
GeV and TeV energies.
We show that the model produces GCR source abundance 
enhancements of the volatile, gas-phase elements, which are an 
increasing function of mass,  as well as a net, 
mass independent, enhancement of  the refractory, grain 
elements over protons, consistent with cosmic ray observations.
We also investigate the implications of the slightly high H/He ratio.
The GCR \nett excess may also be accounted for in terms of the 
acceleration of \netto -enriched pre-SN Wolf-Rayet star wind material 
surrounding the most massive supernovae.
We also show that cosmic ray source spectra, at least below 
$\sim 10^{14}$~eV,
are well matched by the 
model.
\end{abstract}
\clearpage

\section{INTRODUCTION}

The galactic cosmic ray source (GCRS) composition is relatively 
well determined at energies of a few GeV per nucleon.  It is 
taken here to mean the relative values of the differential energy 
fluxes of the various nuclear species, each measured at the 
same energy per nucleon of order a few GeV/$A$ ($A$ is the nuclear 
mass number), and after applying ``standard'' corrections to 
the observed composition for solar modulation and interstellar 
propagation. 
Implicit in this is the assumption that all species have
essentially identical energy spectra at source when plotted as
functions of energy per nucleon, at least in the GeV range
where good composition data can be obtained; observationally this does
seem to be approximately the case
(\egc Swordy 1993; Shibata 1995).  
We note in passing that, whatever the validity of this 
assumption for the nuclear species, cosmic ray 
electrons have very different 
spectra and may, in fact, have entirely different origins (see
Berezinskii et al. 1990 for a detailed account of cosmic ray physics).

If we accept these caveats, the resulting composition data show
very interesting features.
These have been discussed in detail in a companion paper by Meyer, 
Drury, and Ellison (1997; hereafter Paper~I).  We now summarize its 
conclusions.  
As compared to solar photospheric composition, the GCRS composition is 
characterized by a general overabundance of heavier elements relative 
to H and He, and by a fine structure among the heavy elements.
This fine structure is primarily governed by atomic, not nuclear, 
physics; in particular, it does not at all correspond to 
fresh supernova  nucleosynthesis products;
(there exists, however, a \netto, \ct, \oso-rich component in GCRs, 
suggesting the acceleration of some Wolf-Rayet wind material).
The data indicate that the relevant atomic physics parameter could be
either the first ionization potential (FIP), which controls the 
neutral or ionized state of each element in a $\sim 10^{4}$~K gas, or 
the element volatility (i.e., its condensation temperature, \Tco),
which controls the element's ability to 
condense into solid compounds.
For  most elements,  values of FIP and  \Tc are 
anti-correlated, so it is not easy to tell which of these two 
parameters shapes the GCRS composition. In any case,  
{\it either\/} the easily ionized low-FIP elements, {\it or\/} 
the easily condensed high-\Tc elements, are found enhanced by a factor 
of order 5 relative to the other heavy elements (and $\sim $ 30 
relative to H).
 
Most studies to date have considered the FIP hypothesis, largely by
analogy with the situation in the outer solar atmosphere, in which the
solar coronal gas, the solar wind, and the $\sim$MeV solar energetic
particles have undoubtedly a composition biased according to FIP. This
bias implies some ion-neutral fractionation in the underlying cool,
$\sim $ 7000~K chromospheric gas, in which neutrals and ions coexist:
chromospheric ions must rise into the corona more efficiently than
chromospheric neutrals.  If FIP also determines the composition of
GCRs, the cosmic rays must consist of injected $\sim$MeV stellar
energetic particles, originating in F to M later-type stars possessing
a neutral, cool chromosphere similar to that of the Sun, and then
later preferentially reaccelerated to GeV and TeV energies by
supernova shocks (Meyer 1985).

Fortunately, however, a few elements do {\it not\/} fit in the general 
anti-correlation between FIP and \Tco, and these can be used to lift 
the degeneracy and decide which of the two parameters is relevant; 
these are low-FIP volatile elements (especially Na, Ge, Pb) and 
high-FIP refractories (P).  These crucial elements are, unfortunately, 
not among those whose GCRS abundance is easiest to determine!  But, 
with the steady progress over the past years, it has now become 
apparent that all four key ratios Na/Mg, P/S, Ge/Fe, Pb/Pt point 
towards volatility, not FIP, as the relevant parameter.

\placefigure{fig:compmass}

In Paper I, we have shown that all the GCRS composition data are
remarkably well ordered in terms of two specific behaviors for the
volatile and the refractory elements: (i) Among the volatile elements,
the abundance enhancements strongly increase with element mass $A$;
only hydrogen does not entirely fit into the pattern. We believe this
reflects an increase of the acceleration efficiency with the element
mass-to-charge ratio $A/Q$,
\iec  with its rigidity at a given velocity; in any ionization model,
indeed, $A/Q$ is a roughly monotonically increasing function of the
mass $A$ ($Q$ is the charge number).  The low GCRS abundances of H, 
He, and N can be interpreted in this framework.  (ii) The
refractory elements are all enhanced relative to volatiles; but this
enhancement is approximately the
same for all refractories, \iec  it has {\it little or no}
mass-dependence
(see Figure~\ref{fig:compmass}).

This is a very surprising result.  It is quite clear from UV, IR, and
visible observations that the refractory elements are largely locked
into solid dust grains in most of the interstellar medium (ISM) (\egc
Cardelli 1994; Sembach and Savage 1996; Savage and Sembach 1996), as
well as in supernova ejecta (\eg Lucy \etal 1989, 1991; Dwek \etal
1992), and in stellar, and especially Wolf-Rayet star, wind envelopes
(\eg Bode 1988; Gehrz 1991; van der Hucht and Hidayat 1991; van der
Hucht and Williams 1995).  Clearly, the GCRS composition features
suggest a preferential acceleration of those elements locked in grains
in most of the ISM and stellar ejecta (refractories), relative to the
gas-phase elements (volatiles).  This leads to the idea of a
preferential acceleration of grain erosion products in supernova
shocks, an idea which had been earlier approached, along two lines.
Cesarsky and Bibring (1980) and Bibring and Cesarsky (1981)
considered the destruction of grains after their free crossing of the
shock, followed by stochastic acceleration of the grain destruction
products.  Epstein (1980), on the other hand, considered a
preferential acceleration of the entire grains themselves, followed by
their erosion, with the refractory grain erosion products
keeping the high velocity first acquired by the grain (see Paper I for
a fuller discussion).

It is this second line that we want to reexamine in this paper, in the
light of the above, specific composition features, and of modern
nonlinear (i.e., smoothed) shock acceleration theories.  In this
approach, we are encouraged by three basic observations: 
First, if
particles with the same energy per nucleon are considered, smoothed
shock models predict an increased particle acceleration efficiency for
increased particle rigidity, \iec $A/Q$ ratio, because higher rigidity
ions can diffuse further back upstream of the shock than low $A/Q$
particles at the same energy per nucleon.  Hence, high $A/Q$ particles
``see" a larger velocity difference and are more easily injected to
suprathermal energies (\egc Eichler 1979,1984; Ellison and Eichler 1984).
This effect fits qualitatively with the mass dependence of the
volatile, gas-phase element enhancements.
Second, weakly charged grains can behave as extremely high $A/Q$ ions, and
thus get very 
efficiently injected and  accelerated provided they obtain similar
energies per nucleon to protons when first shock heated.
Third, if the refractory elements are, at the early crucial stage,
accelerated as part of grains, their own $A/Q$ plays no role in this
acceleration stage, so that the approximate mass {\it in}dependence of
their GCRS enhancements is not surprising.

In a way, this approach represents a synthesis between earlier fits of
the global enrichment of heavier elements, especially relative to H,
He, but which could not fit the fine structure of the heavy element
composition, such as the Mg/Ne ratio (Ellison 1981;
Ellison, Jones and Eichler 1981; Cesarsky, Rothenflug and Cass\'e
1981), and of approaches which used atomic physics to explain this
fine structure (FIP, or volatility), but could not account for the low
H and He abundances (Meyer 1985).

Leaving the observational study of the GCR composition, theoretical
ideas on particle acceleration in shock waves have been developed to
the point where quite sophisticated models are now routinely
calculated.  In this paper, we calculate expected source composition
and spectral shapes of the GCRs, using a \mc model of cosmic ray
acceleration at SNR shocks (\eg Jones and Ellison 1991; Ellison 1993)
including both interstellar gas and ``grains.''  The shock model
includes nonlinear effects from shock smoothing and a parameterized
description of injection from the thermal background for any ion
species or grain.  It yields both the spectral shapes and absolute
abundances of various species of cosmic rays.  To this we have added a
simple model of grain deceleration, sputtering of individual atoms off
grains, and acceleration of sputtered ions to cosmic ray energies,
yielding a first principles estimate of the refractory
element/hydrogen ratio in cosmic rays.

While protons and helium ions are treated self-consistently and
contribute to the shock smoothing, the other gas-phase ions and grains
are treated as test particles, and are accelerated by the shock as
smoothed by hydrogen and helium.  Despite the approximations that must
be made for such a complex calculation, we find {\it excellent
agreement with observations} for both the spectral shape and the
relative abundances of the various nuclear components, at least above
a few GeV, where solar modulation is no longer important, and below
the observed ``knee" in the GCR spectrum and the Lagage and Cesarsky
(1983) limit at $\sim 10^{14-15}\,\rm eV$ (also Prishchep and Ptuskin,
1981).  We believe this is the first attempt to simultaneously and
self-consistently address the intensity and shape of the major cosmic
ray components using a mixture of interstellar gas and dust.

\section{MODEL ASSUMPTIONS}

The hypothesis we wish to test is that the bulk of the GCRs at
energies below the so-called ``knee'' at $\sim 10^{15}$ eV are
accelerated by the forward shock waves associated with supernova
remnants.  The main arguments in favor of a SNR origin for the GCRs
are that it is very hard to think of any other possible acceleration
sites with adequate power (although stellar winds might be sufficient)
(\eg Axford 1981), and that SNRs are known from their radio
synchrotron emission to accelerate electrons to cosmic ray energies
(\egc Reynolds and Ellison 1992).  We assume the forward shocks in the
Sedov phase are mainly responsible for producing cosmic rays. While
inner reverse shocks exist in the early remnant phases, the forward
shocks are much longer lived and contain more energy. In addition,
cosmic rays accelerated when the remnant is small suffer strong
adiabatic losses making later stages more important(e.g., Drury and
Keane 1995).  The forward shocks accelerate mainly ambient
interstellar material, although recent calculations of the
Raleigh-Taylor instability in {\it young} SNRs (Jun \& Norman 1996)
suggest that some clumps of fast-moving ejecta may actually punch
through the outer shock. However, there are good reasons to believe
that these are relatively minor effects (see Drury and Keane 1995).

In addition, in diffusive shock acceleration (see Drury 1983; V\"olk
1984; Blandford and Eichler 1987; Berezhko and Krymsky 1988; Jones and
Ellison 1991 for reviews), we have a plausible mechanism for producing
energetic particles and one which has been observationally verified to
work at heliospheric shocks under plasma conditions quite similar to
those of interstellar shocks (\egc
\EMP 1990; Baring \etal 1995, although of course at much lower
energies; but this merely reflects the small size and short life-time
of heliospheric shocks).  In its simplest test particle version, this
mechanism predicts that all particles accelerated in a given shock
will have identical power-law spectral shapes in momentum regardless
of their charge (Krymsky 1977; \ALS 1977; Bell 1978; \BO
1978).  The spectra are naturally considered to be momentum spectra,
rather than energy or velocity spectra say, because the basic
acceleration process involves the change in momentum when switching
from the upstream to the downstream reference frame.

However, at a more sophisticated level, where the reaction of the
accelerated particles on the shock structure is considered [\ie the
upstream flow is slowed gradually (smoothed) by the pressure of
backstreaming energetic ions before making an abrupt transition to the
downstream state], the spectra are no longer exact power-laws, and
different ion species are treated differently (\egc Eichler 1984; \EE
1984).  We note that if SNR shocks do indeed power GCRs, the power
required ($\gappeq 10\%$ of the total kinetic energy of SNR ejecta)
implies that reaction effects of the cosmic rays on the shocks must be
important.  These effects produce different spectral shapes (in the
range where all particles are not fully relativistic) and different
injection and acceleration efficiencies for different ion species,
with the ion rigidity and thermal speed
becoming the important distinguishing parameters.
In essence, the question which concerns us in this paper is whether
the differences in the spectral shapes and absolute intensities for
different elements, which inevitably result from this modification of
the shock by reaction effects (Ellison 1993), combined with the
acceleration and destruction of grain material, can explain in a
quantitative way the observed features of the GCR composition.  We
find the answer to be yes, provided that charged grains and gas ions
of the same rigidity act similarly  in the shock acceleration
process.

\subsection{Spectra and Injection in Modified Shocks}

While many different SNR shocks
of varying ages, sizes, and strengths undoubtedly contribute to the
observed cosmic ray flux, we only model single, steady-state
shocks (of varying parameters) and leave more complex models for later
work.

\placefigure{fig:flow}

Let us consider a steady, plane-parallel, modified shock which we take 
to be propagating in the $x$-direction.  The flow velocity profile, 
$\Ux$, will then have a form similar to that indicated 
in Figure~\ref{fig:flow}
by either the solid, dashed, or dot-dashed lines.
At energies (or momenta) high enough 
for the diffusion approximation to be valid, the isotropic part of the 
phase space density, $f_\alpha(x, p),$ for particles of species 
$\alpha$ at position $x$ and momentum $p$ satisfies the well-known 
equation (\egc Drury 1983)
\begin{equation}
U {\partial f_\alpha\over\partial x} =  
{\partial\over\partial
x}\left[\kappa_\alpha(p,x){\partial f_\alpha\over\partial x}\right] +
{1\over 3}{\partial U\over\partial x} p{\partial f_\alpha\over\partial
p} \ ,
 \label{eq:t-eqn}
\end{equation}
where $\kappa_\alpha$ is the $xx$-component of the diffusion tensor
for species $\alpha$ and is expected to be a rather strongly
increasing function of momentum $p$.  At energies close to thermal
energies the diffusion approximation is no longer applicable, however,
the same basic physics can easily be handled by Monte-Carlo
simulations of the particle scattering off magnetic scattering
structures embedded in the plasma flow (\egc
\EE 1984; \EJR 1990; \JE 
1991).  This computational model can be extended down to thermal
energies and used to give a description of the particle distribution
functions through the collisionless shock structure and of the shock
itself.  While it is certainly true that the detailed micro-physics of
the collisionless shock structure is far more complicated than this
simple \mc model, the results do seem to be in good agreement with
heliospheric observations and with more detailed hybrid simulations
(\EMP 1990; \EGBS 1993; Baring \etal 1995, 1997).  Perhaps the best
way to think of it is that this model provides a physically motivated
means of estimating the relative rate at which different species will
be injected and accelerated in a strong collisionless shock, and one
which agrees with such observational evidence as is available for
heliospheric shocks.

If high-energy particles can be removed from the system, either
through a so-called upstream free-escape boundary (FEB), or simply by
escaping when they reach a certain maximum momentum, it is possible to
construct completely steady solutions, regardless of Mach number, in
which the shock modification is self-consistently determined by the
pressure of the accelerated particles.  Such shock structures are
shown in Figure~\ref{fig:flow}. Full details and extensive comparisons with
observations of heliospheric shocks can be found in the papers just
mentioned as well as \EJR (1990) and \EBJ (1995).

One point we wish to make is that, even if the shock structure is
steady and the same for all species, the nonlinear effects of shock
smoothing imply that species with differing diffusion coefficients
will, in general, be accelerated differently and develop different
spectra.  However, we expect that the scattering mean free paths of
the various species will be determined purely by the magnetic rigidity
of each species, and thus, in the relativistic limit, all particles of
a given rigidity will have the same diffusion coefficient.  This
assumes, of course, that other physical constraints, such as the time
available for acceleration being large compared to the particle
gyroperiod, are satisfied.  Spectral differences occur at
nonrelativistic energies because different mass particles of the same
rigidity have different speeds, and this results in a change in
injection efficiency, and the resulting change in absolute accelerated
particle intensities persists into the relativistic regime where the
spectral shapes become the same.\footnote{As we report later in
the paper, our results show that spectral differences between iron and
hydrogen/helium  are noticeable well into the relativistic regime.}
More precisely, we define  the rigidity of species $\alpha$ as
\begin{equation}
R =  {c p\over \Qa e} 
  =  {\mp c^2\over e} \left ( {A \over Q} \right )_{\Z\alpha} 
     \beta \gamma
\end{equation}
(with units of Volts in the SI system), where $c$ is the velocity 
of light {\it in vacuo}, $e$ is the electronic charge, $\mp$ is the 
proton mass, $m_\alpha = \Aa \mp$ is the rest mass of species $\alpha$ 
with $\Aa$ nucleons, $\Qa$ is its charge number, $\beta$ is the 
particle's $v/c$ and $\gamma$ is its Lorentz factor.  The particle 
velocity is then
\begin{equation}
v =  \left [1 + \left(  {m_\alpha c\over p}   \right)^2\right ]^{-1/2}
  = c\left [1 + \left(\Aa \mp c^2\over \Qa e R\right)^2\right ]^{-1/2}
\ .
\end{equation}
In a magnetic field, $B$, the gyroradius of the particle in SI units 
is $\rg = p/(\Qa e B) = R/(c B)$, and we expect from quasi-linear 
theory that the mean free path will be $\lambda(R) \sim \rg/I(R)$, where 
$I(R)$ is the dimensionless power in magnetic field irregularities on 
length scales of order $\rg$ (e.g., Drury 1983).
In all that follows, we assume that $I(R)$ is independent of rigidity
and position relative to the shock and write
$\lambda= \eta \rg$.
The so-called strong scattering Bohm limit obtains when $\eta \sim 1$.

Thus the corresponding spatial 
diffusion coefficient will be
\begin{equation}
\kappa_\alpha = 
{\lambda(R) v\over 3} = 
{c \rg \eta \over 3 }
\left [1 + \left(\Aa \mp c^2\over \Qa e R\right)^2\right ]^{-1/2}
\ ,
\label{eq:kappaeqn}
\end{equation}
which, for non-relativistic velocities, reduces to
\begin{equation}
\kappa_\alpha = 
{\eta \over 3 }  {e \over \mp c^2}  {1 \over B}   
\left ( {Q \over A} \right )_{\Z\alpha}  R ^2   =  
{\eta \over 3 }   {\mp c^2\over e}   {1 \over B}   
\left ( {A \over Q} \right )_{\Z\alpha}  \beta ^2 
\ .
\label{eq:kappaeqnnr}
\end{equation}
We see that particles with the same rigidity but different values of 
$(A/Q)_\alpha$ will have different diffusion coefficients in the 
sub-relativistic region and, therefore, different diffusion lengths in
the shock precursor.
Since the diffusion length in the precursor is 
$L_{\rm D,\alpha} \sim \kappa_{\alpha} / \Ux \propto \L_{\alpha} 
v_{\alpha}$,
the ratio of diffusion lengths of species $\alpha$ to protons, 
at fixed energy per nucleon, is simply
\begin{equation}
{ L_{\rm D,\alpha} \over L_{\rm D,p} } \propto \left ( {A \over Q} 
\right )_{\Z\alpha} 
\ .  
\label{eq:diff-length}
\end{equation}

If the shock is smooth rather than discontinuous (cf.,  Figure~\ref{fig:flow}) 
as will always be the case in nonlinear shock acceleration if $\kappa$ 
is an increasing function of energy, ions with large $A/Q$ will 
diffuse further upstream than protons of the same energy per nucleon 
(provided both are nonrelativistic).  These ions will be turned around 
against a more rapid upstream flow, will receive a larger energy boost 
for each shock crossing, and will be accelerated more efficiently and 
obtain  a 
flatter spectrum than protons (\egc \EJE 1983; \EE 1984).
Recent specific observational support for this effect in anomalous
cosmic rays  has been reported by
Cummings and Stone (1996).
While the differences in $A/Q$ may be small for 
atoms,  grains can have huge $A/Q$'s ($\sim 10^{4-8}$).
If  all particles become 
relativistic, the term in square brackets in Eq.~(\ref{eq:kappaeqn}) 
goes to one and the differences in $\kappa$ vanish, but the
intensities of the spectra of various species, 
acquired  when they were nonrelativistic, will be different, and this 
difference persists to the highest energies obtained.

It is important to note that this conclusion
is based on our assumption of a 
steady state.  The longer diffusion lengths of heavy ions also result 
in longer acceleration {\it times} to a given energy per nucleon.  If 
the shock has a finite age, the acceleration of heavy ions, and
particularly grains, may cutoff before protons.

\subsection{Acceleration of Dust Grains}

We now revive an old suggestion of Epstein (1980) that charged dust
grains could be efficiently accelerated by shocks to produce cosmic
rays (see also Meyer 1993).  A related idea, that grain destruction
products could be stochastically accelerated
behind shocks,  was presented by
Cesarsky and Bibring (1980) and Bibring and Cesarsky (1981).
The basic idea of Epstein (1980)  
was that the dust grains could behave like ions of very large 
mass to charge ratio, thus large rigidity, and should
therefore be relatively 
efficiently accelerated to velocities where they are eroded  by 
sputtering.
The sputtered grain material will have the velocity of the parent 
grain which can be well above 
thermal.  If the sputtering occurs in the upstream region the 
sputtered products (i.e., refractory elements) will be carried back 
into the shock and further accelerated to cosmic ray energies with 
higher net efficiency than the gas-phase thermal protons and 
volatile element ions.
Of course, material which is sputtered from the grains downstream of 
the shock is mainly carried away and lost from the system because the
sputtered ions are, on average, many more mean free paths
downstream from the shock then the parent grain.

Small grains in a plasma will be charged by a number of processes.  In
the absence of any other effects, they acquire a negative charge
because the thermal electron flux impinging on their surface is higher
than the ion flux; this tends to charge the grain to a negative
potential of order $- 2.5 k T/e$ where $T$ is the plasma temperature
(\egc Spitzer 1978).  However a number of competing processes lead to
electron loss from the grain.  A significant UV photon flux can charge
the grain by the photoelectric effect to a positive potential
corresponding to the difference between the photon energy and the work
function of the grain surface.  If the grain is moving through the
plasma, neutral atom and ion impacts on the grain surface can lead to
secondary electron emission which also charges the grain positively.
At $T>10^5{\rm K}$, the charging effect of the thermal electron flux
is largely cancelled by the fact that the electrons themselves produce
secondary electron emission.  All of these processes, together with
poorly known grain properties, make an exact determination of the
grain charge impossible.  What one can say with certainty is that a
grain will only be uncharged very briefly, if at all, and that in
general the potential of the grain will be of order 10 to 100 V (it is
of course no coincidence that this is the characteristic energy range
of atomic physics and electronic transitions).  McKee
\etal (1987) show in their Figure~6 some interesting model calculations 
of grain potentials under various assumptions.

It follows that if the grain potential is $\phi$, the charge on a 
spherical grain is of order $q \sim 4\pi \epsilon_0 a \phi$, where $a$ 
is a characteristic size of the grain (note that the grains will 
certainly not be perfect spheres; in fact the larger grains probably 
have a fractal structure).  In terms of electronic charges this gives 
numerically
\begin{equation}
\QG =  {q\over e} \simeq 700 \left ( {a\over 10^{-7}{\rm m}} \right ) 
\left ( {\phi\over 10{\rm V}} \right ) 
\ . 
\label{eq:charge}
\end{equation}
The number of atoms in the grain will be of order $(a/10^{-10} {\rm 
m})^3$, or $10^9$ for a $10^{-7}{\rm m}$ size grain (or less if the 
grain has a very open fractal structure).  If $\mu$ is the mean 
atomic weight of the grain atoms, the entire grain ``atomic weight,"
 $\AG$,
is $\mu (a/10^{-10} {\rm m})^3$. Thus the effective $A/Q$ value for a 
grain is very large, of order
\begin{equation}
{\left( A\over Q \right)_{\Z\rm G}} \simeq 
1.4\times 10^6 \mu \left(a\over 
10^{-7}{\rm m}\right)^2 \left(\phi\over 10 {\rm V}\right)^{\Z-1}
\ .  
\label{eq:AoverQ}
\end{equation}
If the dust grain has velocity $\betaG c$ with $\betaG \ll 1$, 
then the magnetic rigidity of the grain is
\begin{equation}
R = 
{c p\over q} = 
{\AG \mp \betaG c^2\over \QG e} \simeq 10^9 \betaG 
{\left( A\over Q \right)_{\Z\rm G}}\quad{\rm V}
\ ,
\end{equation}
where $p = \betaG c \AG \mp$ is the momentum of the grain.  
Numerically,
the grain rigidity is
\begin{equation}
R \simeq 
1.3\times10^{15} \betaG \mu \left(a\over 10^{-7}\rm 
m\right)^2 \left(\phi\over \rm 10\, V\right)^{\Z-1} \,\rm V
\ . 
\end{equation}

Ultraviolet and optical extinction measurements indicate that the
grain size distribution is quite broad extending from very small
grains to an upper cutoff at $\sim 0.25 \mu$m (\egc Mathis, Rumpl, and
Nordsieck 1977). The amount of total grain mass in particles with
radii $a$ or less, $\MG(<a)$, goes roughly as $a^{1/2}$, so that
nearly half of the total grain mass is in a relatively small range of
sizes around $0.1\,\mu$m.

In general, supernova remnant shocks have velocities in the range 30
to $3000\,\rm km\,s^{-1}$ (e.g., Reynolds 1988).  Let us consider a
high Mach number shock of velocity $400\,\rm km\, s^{-1}$ which
overtakes a dust grain in a typical interstellar hydrogen density,
$\denH \sim 1$ \pcc.  Relative to the post-shock gas, the grain
will have a velocity of $\sim 300\,\rm km\,s^{-1}$ or $\betaG \sim
10^{-3}$, and thus a rigidity of about $10^{14}\,\rm V$ if it is 0.1
$\mu$m in size, is charged to a surface potential of 10 V, and is made
of material with $\mu\sim 56$.\footnote{While the value $\mu=56$
only applies to pure iron grains, we have chosen it for simplicity.
For silicate grains containing Mg, Si, Fe, and O, $\mu\sim$ 20 to 30,
but this factor of about two difference does not seriously influence
the results that follow.}

Can dust grains be accelerated by SNR shocks?  We assume that SNR 
shocks are capable of accelerating {\it protons} to energies of order 
$10^{14-15}\,\rm eV$, as suggested by theoretical arguments concerning
shock acceleration 
(e.g., Blandford and Eichler 1987),
is required if they are 
to account for the near
constant slope of the observed proton 
spectrum up to the knee at $>\!10^{14}\,\rm eV$,
and will certainly occur if SNRs do, in fact, accelerate 
{\it  electrons} to $\sim10^{14}$ eV, as has been claimed 
for SNR SN
1006 (Koyama \etal 1995; Reynolds 1996).
If this is the case, three conditions must be met; (i) the magnetic 
field near the shock has to contain structures capable of scattering 
protons of rigidities up to $10^{14-15}\,\rm V$, (ii) the shock radius 
(i.e., the size of the acceleration region) must be considerably 
larger than a $\sim 10^{14-15}$ eV proton gyroradius, and (iii) 
the age of the remnant must be greater than the acceleration time to 
$\sim 10^{14-15}$ eV. Our fundamental assumption is: if relativistic 
protons of energy $>10^{14}\,\rm eV$ are being efficiently scattered
and accelerated
(i.e., are being scattered nearly elastically and isotropically in the 
local plasma frame), {\it then so should dust grains with the same 
rigidity}.  There is however one vital difference.  The dust grains, 
far from being relativistic, only have a velocity of order the shock 
velocity, at least initially.  Since we also assume that
the mean free path depends only on rigidity, 
not velocity, the diffusion coefficient of 
the grains is smaller than that of the relativistic protons with the same 
rigidity by a factor of the grain $\beta$, typically $10^{-3}$.  

For diffusive shock acceleration, the standard estimate of the 
acceleration time, $\acctime$, to a momentum $p$ is
\begin{equation}
\acctime = {3 \over \Delta U} \int^{\rm p}_{\rm p_i} \left ( {\kappa_1 
\over U_1} + {\kappa_2 \over U_2} \right ) {dp' \over p'}
\ ,
\label{eq:acctimediff}
\end{equation}
where $\kappa_1$ ($\kappa_2$) is the upstream (downstream) diffusion 
coefficient parallel to the shock normal, $U_1$ ($U_2$) is the far 
upstream (downstream) flow speed measured in the shock frame, $\Delta 
U = U_1 - U_2$, and $p_{\rm i} \ll p$ is the injection momentum of the 
particle (\egc Axford 1981).  Using this, we can define the
instantaneous acceleration time scale, $\accscale= p/(dp/dt)$, as
\begin{equation}
\accscale \simeq {3 \over \Delta U} {\kappa_1 \over \Vsk} (1 + gr)
\ , 
\label{eq:acctimescale}
\end{equation}
where $r= U_1/U_2$ is the shock compression ratio, $\Vsk = U_1$ is the 
shock speed, and we have taken $\kappa_2 = g \kappa_1$.  The parameter 
$g$ is expected to lie between 0 and 1 and will equal $1/r$ if the
diffusion coefficient is inversely proportional to background 
density. We
consider $g=1/r$ a good approximation (see \EMP 1990)
and will use this 
in the derivation which follows.  We will also approximate
$\Delta U = \Vsk (1 - 1/r) \sim \Vsk$ in the following expressions.

Therefore, using  $\lambda = \eta \rg$ and Eq.~(\ref{eq:AoverQ}), 
the acceleration time scale of a nonrelativistic
grain can be written as 
\begin{equation}
\accscale \simeq 
10^4
\eta
\left ( {\mu \over 56} \right )
\left ( {a \over 10^{-7} {\rm m}} \right )^{\Z2} 
\left ( {\phi\over 10{\rm V}} \right )^{\Z-1} 
\left ( {B \over 3 \mu{\rm G}} \right )^{\Z-1} 
\left ( {\Vsk \over 400 \> { {\rm km} \over {\rm s}} } \right )^{\Z-2}
\left ( {\betaG \over 0.01} \right )^{\Z2} 
\ {\rm yr} 
\ .
\label{eq:acctimescaletwo}
\end{equation}
Initially, the acceleration rate for grains with $\vG \sim \Vsk =
400$ \kmps \ 
(i.e., $\betaG \sim 0.0013$), is very fast, i.e., $\accscale \simeq 
200$ yrs  for our standard grain parameters and assuming $\eta=1$.  
Strictly speaking, 
Eq.~(\ref{eq:acctimediff}) (which is based on diffusive acceleration theory)
is inapplicable in this low velocity 
limit, but Monte-Carlo simulations (\egc \EBJ 1995)
have shown that it is  quite
accurate down to thermal energies.  We have been working with the momentum 
acceleration time-scale, $p/(dp/dt)$; the kinetic energy acceleration 
time-scale, $E/(dE/dt)$, will be exactly half this, as long as the 
grains are non-relativistic and $E \propto p^2$.

An important constraint is that the age of the remnant, $\tSNR$, is
greater than the acceleration time and that the acceleration time is
greater than the gyroperiod of the grain, $\TG=2 \pi \rg/\vG$, 
so that the grain has
time to spiral and scatter in the background magnetic field, i.e.,
\begin{equation}
\TG \ll \accscale < \tSNR
\ .
\label{eq:time-constr}
\end{equation}
The gyroperiod of a nonrelativistic grain is
\begin{equation}
\TG  \simeq 
10 
\mu  
\left ( a\over 10^{-7}\,\rm m \right )^{\Z2}
\left ( \phi\over 10\,\rm V \right )^{\Z-1}
\left ( B\over 3\,\rm \mu G \right )^{\Z-1}
\ {\rm yr}
\ ,
\end{equation}
and
\begin{equation}
{\accscale \over \TG} \simeq
20
\eta
\left ( {\betaG \over 0.01} \right )^{\Z2}
\left ( {\Vsk \over 400 \> { {\rm km} \over {\rm s}} } \right )^{\Z-2}
\ ,
\end{equation}
so the left-half of eq.~(\ref{eq:time-constr}) is easily satisfied for
$\Vsk\sim 400$ \kmps, $\betaG=0.01$,  and $\eta=1$.
Therefore, it is
reasonable to assume (as we do) that grains can interact
collisionlessly by gyrating in the turbulent magnetic fields.
At the beginning of the acceleration process,
 $\accscale/ \TG$ can certainly be less than one
but eq.~(\ref{eq:acctimediff}) doesn't describe 
the acceleration rate 
in the first few crossings of the shock.

For the right-half of eq.~(\ref{eq:time-constr}), we use the 
standard Sedov solution (e.g., Lang 1980)
for an explosion of initial
energy, $\EnSN$, in a gas of density $\rho= 1.4 \dengas \mp$, 
to estimate $\Vsk$ as a function of $\tSNR$, i.e.,
\begin{equation}
\Vsk =
{2 \over 5}
\xi
\left ( { \EnSN \over \rho } \right )^{\Z1/5}
\tSNR^{-3/5}
\ ,
\end{equation}
where $\xi \sim 1.15$ (\egc Shu 1992) and $\dengas$ is the hydrogen number
density.  Using this value of $\Vsk$ in eq.~(\ref{eq:acctimescaletwo})
we have
\begin{eqnarray}
   & & { \tSNR \over \accscale } \simeq
{ 2 \over \eta }
\left ( { \mu \over 56} \right )^{\Z-1}
\left ( a\over 10^{-7}\,\rm m \right )^{\Z-2}
\left ( \phi\over 10\,\rm V \right )
\left ( B\over 3\,\rm \mu G \right )
\left ( {\betaG \over 0.01} \right )^{\Z-2}  \ \times 
\qquad\qquad\qquad\qquad \nonumber\\ 
   & & \qquad\qquad\qquad\qquad\qquad\qquad\qquad 
\left ( \dengas \over 1\,\rm cm^{-3} \right )^{\Z-2/5} 
\left ( { \EnSN \over 10^{51} {\rm erg}} \right )^{\Z2/5}
\left ( {\tSNR \over 10^3 {\rm yr}} \right )^{\Z-1/5}
\ ,
\label{eq:enflux}
\end{eqnarray}
demonstrating that the right-half of eq.~(\ref{eq:time-constr}) can also
be satisfied.

An additional constraint for acceleration is that 
the diffusion length  must be less than
the shock radius, $\Rsk$, i.e.,
\begin{equation}
{ \kappa \over \Vsk  \Rsk } < 1
\ ,
\end{equation}
or, using
\begin{equation}
\Rsk =
\xi
\left ( { \EnSN \over \rho } \right )^{\Z1/5}
\tSNR^{2/5}
\end{equation}
from the Sedov solution,
\begin{eqnarray}
   & & 
0.03
\eta
\left ( { \mu \over 56} \right )
\left ( a\over 10^{-7}\,\rm m \right )^{\Z2}
\left ( \phi\over 10\,\rm V \right )^{\Z-1}
\left ( B\over 3\,\rm \mu G \right )^{\Z-1}
\left ( {\betaG \over 0.01} \right )^{\Z2} 
\left ( \dengas \over 1\,\rm cm^{-3} \right )^{\Z2/5} \ \times
\qquad\qquad\qquad\qquad \nonumber\\ 
   & & \qquad\qquad\qquad\qquad\qquad\qquad\qquad 
\left ( { \EnSN \over 10^{51} {\rm erg}} \right )^{\Z-2/5}
\left ( {\tSNR \over 10^3 {\rm yr}} \right )^{\Z1/5}
< 1
\ .
\end{eqnarray}
This is easily satisfied for the values of $\betaG$ we consider.

\subsection{Loss Time Scales}

The acceleration time-scale from shock acceleration has to be compared 
to the momentum loss time-scale due to frictional coupling of the 
grain to the background plasma to determine whether acceleration 
really occurs.  There are two components to the frictional 
interaction.  First, there is a component due to direct collisions 
of the grain with atoms or ions of the plasma.  Since even for a 
$10^6$~K gas, $\vG \geq \Vsk > v_{\rm H, thermal}$, the collision rate is 
of order $\dengas a^2 \betaG c$, where each collision reduces the grain 
momentum by a fraction of order $1.4/\AG$ (assuming that collisions with 
hydrogen dominate and that helium increases the loss rate by about 
40\%).  Grain-grain collisions may also be important but this process 
depends on the distribution of grain sizes, is highly uncertain, and
is not expected to contribute much to momentum losses since grains
contain a small fraction of the total mass of the ambient gas 
(K. Borkowski, private communication).   
Thus, neglecting grain-grain collisions, the momentum 
loss time-scale resulting from direct collisions, $\lossmom$, is of 
order
\begin{equation}
\lossmom \simeq { \AG \over 1.4 \dengas a^2 \betaG c} \simeq 8 \mu 
\left(a\over 10^{-7}\rm m\right) \left(\dengas \over 1\,\rm 
cm^{-3}\right)^{-1} \betaG^{-1} \ {\rm yr}
\ ,
\label{eq:timedirect}
\end{equation}
and the ratio of this loss time-scale to the acceleration time-scale 
is
\begin{equation}
{\lossmom \over \accscale} \simeq
{4\x{-6} \over  \eta }
\left ( a \over 10^{-7} \rm m \right )^{\Z-1} 
\left ( \dengas \over 1\,\rm cm^{-3} \right )^{\Z-1} 
\left ( \phi \over \rm 10\,V \right ) 
\left ( B\over 3\,\rm \mu G \right ) 
\left ( {\Vsk \over 400 \> { {\rm km} \over {\rm s}} } \right )^{\Z2}
\betaG^{-3} 
\ ,
\label{eq:lossdirOacc}
\end{equation}
interestingly with no dependence on grain composition, $\mu$.

Secondly, we have to consider the indirect drag on the grain resulting 
from long-range electrostatic interactions.  
It is easy to show (Draine and Salpeter, 1979) that these effects are less
important than direct collisions and we neglect them in what follows.
Therefore, from 
equation~(\ref{eq:lossdirOacc}) we conclude that losses become progressively 
more important as the grains are accelerated to higher velocities.  
This will quench the acceleration at the velocity where $\accscale 
\sim \lossmom$, i.e.,
\begin{equation}
\betaGmax \simeq 
0.016 
\eta^{-1/3} 
\left ( {\Vsk \over 400 \> { {\rm km} \over {\rm s}} } \right )^{\Z2/3}
\left ( a\over 10^{-7}\,\rm m \right )^{\Z-1/3} 
\left ( \dengas \over 1\,\rm cm^{-3} \right )^{\Z-1/3}
\left ( \phi\over 10\,\rm V\right )^{\Z1/3}
\left ( B \over 3\,\rm \mu G\right)^{\Z1/3}
\ ,
\label{eq:betamax}
\end{equation}
or
\begin{equation}
\betaGmax \simeq 0.016
\end{equation}
for our standard  parameters: 
$\Vsk=  400$ \kmps,
$\eta =  1$,
$a= 10^{-7}$ m,
$\dengas= 1$ \pcc,
$\phi= 10$ V,
and $B= 3$ $\mu$G.
So, the grain $\betaG$ can be increased from an 
initial value of order $10^{-3}$ to one of order $10^{-2}$,
with a rather weak dependence on the ambient density, magnetic field, 
and grain size.

This can be written in terms of the maximum energy per nucleon, i.e.,
\begin{eqnarray}
   & & 
\left ( {E \over A} \right )_{\Z\rm G,max}
\simeq 
100 
\eta^{-2/3} 
\left ( {\Vsk \over 400 \> { {\rm km} \over {\rm s}} } \right )^{\Z4/3}
\left ( a\over 10^{-7}\,\rm m \right )^{\Z-2/3}
\left ( \dengas \over 1\,\rm cm^{-3} \right )^{\Z-2/3} \ \times
\qquad\qquad\qquad\qquad \nonumber\\ 
   & & \qquad\qquad\qquad\qquad\qquad\qquad\qquad 
\left ( \phi\over 10\,\rm V \right )^{\Z2/3}
\left ( B\over 3\,\rm \mu G \right )^{\Z2/3}
\ {\rm keV}
\ ,
\label{eq:energymax}
\end{eqnarray}
So  our standard parameters yield $(E/A)_{\rm G,max} \simeq 100$ keV, 
well above thermal energies.  Of course this does not mean that all 
grains are accelerated by this amount, in fact, a distribution
extending upwards from thermal energies will result with only a small
fraction of grains obtaining the  cutoff energy $(E/A)_{\rm G,max}$.  
In addition, as described above, the shock must be large 
enough and old enough for acceleration to these velocities to occur.

\subsection{Grain Sputtering and Injection of Sputtered Material}

The acceleration of grains has significant implications for grain 
erosion  by sputtering.  At the velocities indicated in 
equation~(\ref{eq:betamax}), the sputtering process is quite uncertain and
grains may even become transparent (\egc Dwek 1987),  but we
assume that roughly 0.5 to 1\% of collisions with 
ambient gas atoms  result in the sputtering of an atom from the 
grain surface.  On average, 
each such event reduces $\AG$ by $\mu$.  Therefore, the grain 
destruction time-scale for collisional sputtering, $\losssput = 
\AG/(d\AG/dt)$, is approximately
\begin{equation}
\losssput \simeq { 100 \AG \over \dengas a^2 \betaG c \mu} 
\ .
\label{eq:timesputter}
\end{equation}
Comparing $\losssput$ to the momentum-loss time-scale 
$\lossmom$ for direct collisions with the same gas atoms 
(eq.~\ref{eq:timedirect}), we get
\begin{equation}
{\losssput \over \lossmom} \simeq {140 \over \mu} 
\ ,
 \label{eq:losssputOlossdir}
\end{equation}
so that $\losssput \sim O(10) \times \lossmom$ for $\mu\sim 20$.
For instance, the larger grains ($a \sim 10^{-7}$~m) 
accelerated to $\betaG \sim 10^{-2}$, which have loss and 
acceleration time-scales of order $10^4\,\rm yr$  for 
$\dengas=1~\rm cm^{-3}$, have sputtering time scales of some 
$10^5\,\rm yr$.
The key point is that the acceleration time-scales are always shorter 
than the destruction time-scales.  This means that the grain has time 
to diffuse back and forth between both sides of the shock before being 
destroyed.  Thus, some of the sputtering must occur while the 
grain is in the upstream region ahead of the shock.  These sputtered 
particles can then be advected into the shock as a seed population of 
energetic ions which can then be accelerated with high 
efficiency, if they do not lose too much energy to ionization and 
Coulomb collision losses while being advected from the sputtering site 
to the shock.

The accelerated dust grains will be distributed in the upstream region
with a roughly exponential spatial distribution falling on a
length-scale of order $\kappa/\Vsk$. The time-scale for the sputtered
ion to be advected back to the shock, $t_{\rm ad}$, is thus
$\kappa/\Vsk^2$ (note that the much smaller rigidity of the ion
compared to the parent grain means that we can ignore the diffusion of
the ion) and is thus of order $t_{\rm acc}/6$
(\ie equation~\ref{eq:acctimescale}). If we consider grains
accelerated to the point where the acceleration just balances the
losses due to collisions  (Eq.~\ref{eq:betamax}), which is where
most of the sputtering occurs, the advection time-scale of the
sputtered ions is
\begin{equation}
t_{\rm ad} \simeq
1.6\x{3}
\eta
\left ( {\mu \over 56} \right ) 
\left ( {\betaG \over 0.01} \right )^{\Z2}
\left(a \over 10^{-7}\,\rm m\right)^{\!2}
\left(\phi \over 10\,\rm V\right)^{\Z-1}
\left(B\over 3\,\rm \mu G\right)^{\Z-1}
\left ( {\Vsk \over 400 \> { {\rm km} \over {\rm s}} } \right )^{\Z-2}
{\rm yr} ,
\label{eq:tad}
\end{equation}
typically $\sim 2000$ years for the larger, $a \sim 10^{-7}$~m, 
grains, but falling as the second power of the grain size for smaller 
grains.  The corresponding advection length scale,
$\Vsk t_{\rm ad}$, is of order of a parsec.

We must now estimate the losses of the sputtered ions on the advection 
time-scale.  The ion is unlikely to be fully stripped, especially 
if of high nuclear charge.  On ejection from the grain it will carry 
some electrons with it, and there will also be electron exchange with 
the atoms of the background plasma.  The cross section for electron 
stripping or pick-up is of order $10^{-16}\,\rm cm^{2}$, so that in a 
medium of hydrogen number density $\dengas \sim 1\,\rm cm^{-3}$ and at a 
velocity of order $\betaG \sim 10^{-2}$, the time-scale for electron 
exchange is of order $\sim 1$~yr.  
Electrons will be stripped or picked-up until the kinetic energy of 
colliding electrons in the ion's frame is of the same order as 
the ion's ionization potential.    
In the case of a cold plasma, in which $\beta_{\rm e,th} \ll \betaG$ 
for virtually all thermal electrons in the Maxwell tail, this 
corresponds to ionization potentials  
of $\sim 25 (\betaG/10^{-2})^2$~eV. For $\betaG 
\sim 10^{-2}$, this is of order the 1$^{\rm st}$, 2$^{\rm nd}$, or at 
most 3$^{\rm rd}$ ionization potential 
of all elements.  Thus we expect the ions to have 
an effective charge $Q^*$ of at most $+3$
and those elements with high first ionization potentials actually have 
a significant chance of becoming neutral atoms.
Such neutrals, if formed, are no longer trapped by 
the magnetic field and move in a straight line until they again become 
ionized.  If the charge exchange time-scale is of order one year, the 
displacement through this effect is of order $10^{14}\,\rm m$, 
substantially less than the size of the advection region except for 
the very smallest grains where it may be a significant effect.

The above estimates apply to a 10$^{4}$~K plasma, in the absence of 
steady photoionization.
In a hot plasma consisting of ions with 
ionization potentials $ > 25$~eV, 
the equilibrium charge is determined, not by the accelerated 
ion velocity $\betaG$, but by the thermal balance of the gas.  This 
applies to a 10$^{6}$~K plasma, in which, e.g., accelerated Mg and Fe 
ions will reach effective charges of $Q^* \sim$ +8 and +9, respectively.

The momentum loss time-scale of an ion  with velocity 
$\betaGmax$, due to ionization energy loss and Coulomb collisions is
\begin{equation}
t_{\rm ion, loss} \sim 
2000 {\mu\over Q^{*2}}
\left(\betaGmax \over 10^{-2}\right)^{\Z3}
\left(n_e\over 1\,\rm cm^{-3}\right)^{\Z-1}
({\rm ln} \Lambda)^{-1} 
\Psi^{-1} 
\ {\rm yr} 
\ ,
\end{equation}
in which $\Psi (\vGmax/v_{\rm e,th})$ describes the 
decrease of the momentum loss rate in hotter plasmas, in which the 
thermal electron velocity $v_{\rm e,th}$ becomes comparable to the 
energetic ion velocity $\vGmax$ (Spitzer 1962; Ryter et al.\ 1970).  
With $\ln{\Lambda} = O(20)$, $\mu= O$(50), $Q^{*2}= O(4)$, 
$\Psi=1$ for $T=10^{4}$~K, and $Q^{*2}= O(100)$, $\Psi=0.16$ for 
$T=10^{6}$~K, we get $t_{\rm ion, loss} \sim 1000$ and $\sim 300$~yr 
for $10^{4}$~K and $10^{6}$~K gases.
These figures are roughly comparable to, but somewhat smaller than the 
$\sim$~2000~yr advection time-scale we estimated for the larger 
grains, so that energy loss can be significant, especially for a hot 
ambient gas. 
However, the sputtering occurs throughout an extended region.  This 
means that even for the largest grains, which have the most spatially 
extended distribution, those ions sputtered near the shock are 
advected into the shock before they have suffered significant energy 
losses.  The surviving fraction can be estimated  as 
$1-\exp(-t_{\rm ion,loss}/t_{\rm ad}) \sim 1/2.5$ for 
$T=10^{4}$~K and $\sim 1/7$ for $T=10^{6}$~K for the largest 
grains, but it rapidly tends to 1 for smaller grains (Eq.~\ref{eq:tad}).  
The actual importance of this loss, of course, also depends on the gas 
density.
In the subsequent rough approach, we will neglect this loss, and 
assume that all of the sputtered ions are convected back to the shock.

\subsubsection{Refractory Element  Injection Rate}

We are now in a position to estimate the suprathermal refractory 
element injection rate resulting from the sputtering of grains 
in the upstream region, followed by the advection of the resulting 
suprathermal ions into the shock.  If the number density of thermal
grains far 
upstream of the shock is $\denGr$, 
and the velocity spectrum of the accelerated grains at and 
downstream of the shock is of the form 
$N(\vG)= N_0 (\vG/v_0)^{-2}$ from 
an initial velocity $v_0\sim 3U/4$ to a maximum velocity 
$\vGmax \sim 10v_0$, then
\begin{equation}
\int_{v_0}^{\vGmax} N_0\left(\vG\over v_0\right)^{\Z-2}d\vG \sim
4 \denGr
\ ,
\end{equation}
or $v_0 N_0\sim 4 n_{\rm G}$.  Note that we have assumed that the 
shock has a compression $r= 4$ and the accelerated grain spectrum 
has the corresponding test-particle power law with index $-2$.  
While we have included all downstream grains (even
thermal ones) in the power law spectrum, and our parameters, as well as
the shape of the spectrum, will be modified 
substantially in our nonlinear models presented below, the following
calculation is useful as a rough  estimate of the injection 
efficiency.

To calculate the injection rate, we assume that the sputtering rate 
per grain is of order $0.01 \dengas a^2 \vG$, and that the grains are 
exponentially distributed in the upstream region on a scale 
$\kappa(\vG)/\Vsk$.  Further assuming that the sputtered 
ions have the same velocity, $\vG$, as the parent grain, we find the ion 
injection rate per unit surface area of the shock for ions with 
velocities between $v$ and $v+dv$ is
\begin{equation}
I(v) dv \simeq
{ {\rm sputters} \over {\rm unit \ time \cdot grain} } \times
{ {\rm grains \ in} \ dv \over {\rm volume} } \times
{ {\rm scale \ length} } \times dv
\ ,
\label{eq:injwords}
\end{equation}
that is,
\begin{equation}
I(v) dv \simeq
[0.01 \dengas a^2 v] 
\left [ N_0\left(v\over v_0\right)^{\Z-2} \right ] 
\left [ {\kappa(v)\over \Vsk} \right ] dv ,
\qquad v_0<v<10 v_0
\ ,
\label{eq:injkappa}
\end{equation}
where we have dropped the subscript `G' on velocity.
Replacing $N_0$ with $4 \denGr/v_0$, noting that 
$\kappa(v)/\Vsk^2\simeq \accscale/6$, and that $\dengas a^2 v \simeq 
\AG/(1.4 \lossmom)$ 
(Eqs.~\ref{eq:acctimescale} \ and \ref{eq:timedirect}), we 
obtain
\begin{equation}
I(v) dv \sim
5\x{-3} \denGr \AG {\Vsk \over v_0}
\left ({v \over v_0} \right )^{\Z-2}
{ \accscale \over \lossmom} dv
\ .
\end{equation}

Now, from equation~(\ref{eq:injkappa}), we see that, since $\kappa\propto v^2$ 
(Eq.~\ref{eq:kappaeqnnr}), the injection rate rises as $I(v)\propto 
v$.  The maximum value is obtained at 
$\vGmax \sim O(10 v_0)$, where the 
acceleration time scale and the direct collisional momentum loss 
time-scales coincide, i.e., where $\accscale \sim \lossmom$.  This 
implies that
\begin{equation}
I(v) \sim 
5\x{-3} \denGr \AG { \Vsk \over v_0 }
\left ( {\vGmax \over v_0} \right )^{\Z-2} 
{v \over \vGmax }
\ ,
\end{equation}
and thus the total injection rate of sputtered ions per
unit surface area is
\begin{equation}
\qsput =
\int_{v_0}^{\vGmax} {I(v)} dv ~ \sim ~
5\x{-3} \denGr \AG { \Vsk \over v_0 }
\left ( {\vGmax \over v_0} \right )^{\Z-2} 
{ \vGmax^2 - v_0^2 \over 2 \vGmax }
\ ,
\end{equation}
or
\begin{equation}
\qsput \sim O(10^{-4}) \denGr \AG \Vsk
\ .
\label{eq:inj-estimate}
\end{equation}

The quantity $n_{\rm G} \AG \Vsk$ is the flux of nucleons 
contained in grain 
material coming from far upstream.  This result can be interpreted as 
saying that, with a probability of order $10^{-4}$, an atom in a dust 
grain will be sputtered as a suprathermal ion while in the upstream 
region  and be convected back to the shock without major energy 
loss (at least in a $10^4$ K gas).
The resulting suprathermal ions typically 
have velocities of order 10 times the downstream thermal proton 
velocity and, in a $10^{4}$~K gas without 
ongoing photoionization,
have a charge of $Q^* \leq +3$.
In a $10^6$ K gas, the mean charge can be higher 
(\ie $\sim\,+9$ for Fe) and energy losses for the sputtered ions may
be significant.

It is instructive to examine where the factor of $10^{-4}$ comes from; 
two of the four decades come from the low sputtering yield per 
collision, one from the fact that the acceleration time scale is 
typically a decade longer than the time-scale for advection out of the 
upstream region, and one from the fact that the number density of the 
accelerated grains per logarithmic interval decreases as $v$,
while $\vGmax$ is about a decade above $v_0$.  
We note in passing that this relatively low 
probability also means that the process of grain acceleration does not 
significantly change discussions of grain destruction and processing 
by shock waves, at least in the upstream region.

In view of the rather crude nature of this estimate it is remarkable 
that the answer appears close to what is required by the GCR 
composition observations. As is well known, if one simply assumes 
that the accelerated protons have a $p^{-2}$ power-law spectrum from a 
few times thermal energy to an upper cut-off at around $10^{14}\,\rm 
eV$, the condition that the total energy flux in accelerated protons 
out of the shock cannot exceed the mechanical power in,
implies that only about 1 in $10^4$ of the incident 
thermal protons can become part of the  cosmic ray proton power law.
The coincidence of this figure with the estimate 
for the ion sputtering probability suggests that the resultant 
accelerated cosmic ray composition will be fairly close to the average 
chemical composition of the interstellar medium.  But, since 
the sputtered ions are injected at a velocity 
about a decade higher than the protons, the refractory 
elements should show an enhancement which is also of order 10 (see
Fig. 3.4 in Jones and Ellison 1991).

\subsection{Detailed Assumptions of \mc Shock Model}

In the \mc technique used here, we model a plane, parallel,
steady-state shock, \iec the angle between the upstream magnetic field
and the shock normal, $\Tbn$, is assumed to be zero everywhere and the
shock is taken to be an infinite plane.  We mimic the curvature of a
real SNR shock by placing a free escape boundary (FEB) at some
distance, $\dFEB$, upstream from the shock.  Shocked particles
reaching the FEB are lost from the system, thus truncating the
acceleration process.  We note that in the calculation of Berezhko,
Elshin, and Ksenofontov (1996), the size of the SNR shock, rather than
its age, also limits the acceleration process. In a steady state, with
a diffusion coefficient which increases with energy, a FEB boundary is
required to obtain self-consistent solutions in all but extremely low
Mach number shocks (Eichler 1984; \EE 1984).  
%
% zzz - 3 
%
It has been shown in Kang and Jones (1995) (see also Knerr, Jokipii,
and Ellison 1996) that a FEB works in a similar fashion in
time-dependent, plane shocks.
%
% zzz - 3
%
Our steady-state assumption precludes a description of the overall
dynamics of the SNR explosion; instead, we use standard Sedov estimates for
SNR shock radii, speeds, and ages where these are required.

The \mc model makes the same assumption for the scattering mean free
path as made to obtain equation~(\ref{eq:acctimescaletwo}), \iec
\begin{equation}
\L =  \eta \rg
\ ,
\label{eq:mfp-eqn}
\end{equation}
where $\eta$ is a constant independent of particle species, energy, or 
position. The diffusion 
coefficient is then $\L v/3$, where $v$ is the particle speed in the 
local plasma frame.  Henceforth, all lengths will be measured in units 
of $\Lz= \eta\rgone$, where $\rgone= \mp \Vsk /(eB_1)$ is the 
gyroradius of a far upstream proton with a speed equal to the shock 
speed.  We further assume that all particles scatter elastically and 
isotropically in the local plasma frame.  By assuming that the 
scattering is elastic against a massive background, we model a 
situation where particles scatter against waves which are frozen-in 
the plasma.  This assumption ignores the possible transfer of energy 
between the particles and the background wave field,
and will be most accurate for high 
Mach number shocks.  Obviously, the assumption that 
equation~(\ref{eq:mfp-eqn}) holds over many orders of magnitude from thermal 
energies to $10^{14-15}$ eV is a gross simplification of the complex 
plasma physics which controls particle diffusion in the self-generated 
magnetic turbulence near shocks.  However, $\rg$ is the fundamental 
length scale for scattering, and equation~(\ref{eq:mfp-eqn}) does model 
strongly energy dependent diffusion.  In addition, equation~(\ref{eq:mfp-eqn}) 
has been shown to (a) be consistent with spacecraft observations of 
protons and heavy elements accelerated at the quasi-parallel Earth bow 
shock (\EMP 1990), (b) allow models of Ulysses spacecraft data of 
protons and \Hetwo accelerated at interplanetary traveling shocks 
(Baring \etal 1995), (c) be of a similar form to that determined 
directly from self-consistent plasma simulations (Giacalone \etal 
1992,1993), and (d) match plasma simulation results for injection and 
acceleration when used in the \mc simulation we employ here (\EGBS 
1993).  It must be cautioned, however, that all of the above 
comparisons were performed in energy ranges far smaller than
those modeled here; nevertheless we feel this expression for the 
diffusion contains the essential physics of the processes involved and 
allows reasonably self-consistent models, which can be meaningfully 
compared to observations.

We further assume the SNR shocks in question are capable of
accelerating {\it protons} to energies on the order of $\Epmax \sim
10^{14-15}$ eV. This limit is imposed by the observed constancy of the
energetic proton spectral shape up to those energies (Shibata 1995);
cutoffs above this energy can be interpreted in terms of either the
finite size of the shock acceleration region (\eg Berezhko, Elshin,
and Ksenofontov 1996), or the finite age of the remnant (\eg Prishchep
and Ptuskin, 1981; Lagage \& Cesarsky 1983), depending on the
parameters.  For our models here,
% zzz
we assume 
a finite shock size limits proton acceleration.
If the waves
responsible for scattering high energy particles are self-generated,
the upstream diffusion length of the highest energy particles
currently in the system will define the turbulent foreshock region.
Energetic particles backstreaming to the limits of the foreshock
region will leave the system truncating the acceleration.
Since we assume that the magnetic turbulence responsible for 
isotropizing protons of a given gyroradius will act similarly on other 
species (including grains!)  of the same gyroradius, the acceleration 
of all other species will cease when their diffusion lengths in 
the upstream medium, $L_{\rm D,\alpha} = \kappa_{\alpha}/\Vsk$, 
equals that of the highest energy protons.
Since $\kappa_{\alpha} = \lambda_{\alpha} v_{\alpha}/3$ and 
$\lambda_{\alpha} = \eta r_{\rm g,\alpha}$,
the shocks will be able to 
accelerate an ion of species $\alpha$ up to an energy 
such that
\begin{equation}
v_{\alpha} p_{\alpha} =  \Epmax  Q_{\alpha} 
\ ,
\end{equation}
or
\begin{equation}
\left ( {E\over A} \right )_{\Z\rm \alpha,max} =  \zeta  
\left ( {Q\over A} \right )_{\Z\rm \alpha} \Epmax
\ ,
\label{eq:EAmax}
\end{equation}
where $(E/A)_{\rm \alpha,max}$ is the approximate maximum kinetic 
energy per nucleon a species $\alpha$ will obtain in a shock large 
enough to accelerate protons to $E_{\rm p,max}$.  
Here $\zeta$ = 1 
for highly relativistic maximal energies ($\alpha$ = true ion, 
$(Q/A)_{\alpha} > 10^{-2}$) and $\zeta$ = 1/2 for non-relativistic ones 
($\alpha$ = grain, $(Q/A)_{\rm G} \sim 10^{-8}$).  For $\Epmax \geq 
10^{14}$~eV, we get for grains $(E/A)_{\rm G,max} \geq 500$~keV, due 
to the finite size of the SNR.
As seen above, the energies actually reached by grains are limited to 
only $(E/A)_G \lappeq 100$~keV by collisional friction ({\S}~2.3).

In terms of the actual SNR environment, the maximum energy 
depends on three parameters, $\dFEB$ which
is some measure of the shock radius, 
$\eta$, and the magnitude of the upstream magnetic field, $B_1$.
As long as we confine ourselves to 
parallel shocks, these three parameters can be combined into one.
We take $\dFEB$ to be some fraction $f$ of the shock radius 
$\Rsk$, and set this distance equal to the upstream 
diffusion length, \iec $f \Rsk = \kappa_1/\Vsk$.
For highly relativistic particles, this gives:
\begin{equation}
{ \eta \rg c \over 3 \Vsk } = f \Rsk
\ , 
\end{equation}
and since
\begin{equation}
\rg =  { A \over Q } {(E/A) \over e B_1 c} 
\ ,
\quad\quad {\rm (SI \ units)}
\end{equation}
we obtain for the maximum kinetic energy per nucleon,
\begin{equation}
\left ( { E \over A } \right )_{\Z\rm max} =  
{ Q \over A } { \left ( f B_1 \over \eta \right ) }
 3 e
\Rsk \Vsk 
\ .
\end{equation}
Replacing  $\Rsk$ and $\Vsk$ with their Sedov values at $\tSNR$ we
have
\begin{equation}
\left ( { E \over A } \right )_{\Z\rm max} \simeq
3\x{14}
\left ( {Q \over A} \right )
\left ( {f B_1 \over \eta \ 3\mu {\rm G}} \right )
\left ( \dengas \over 1\,\rm cm^{-3} \right )^{\Z-1} 
\left ( { \EnSN \over 10^{51} {\rm erg}} \right )^{\Z2/5}
\left ( {\tSNR \over 10^3 {\rm yr}} \right )^{\Z-1/5}
\ {\rm eV}
\ ,
\end{equation}
or, in terms of $\Vsk$,
\begin{equation}
\left ( { E \over A } \right )_{\Z\rm max} \simeq
2\x{14}
\left ( {Q \over A} \right )
\left ( {f B_1 \over \eta \ 3\mu {\rm G}} \right )
\left ( \dengas \over 1\,\rm cm^{-3} \right )^{\Z-1/3} 
\left ( { \EnSN \over 10^{51} {\rm erg}} \right )^{\Z1/3}
\left ( { \Vsk \over 10^3 {\rm { km \over s} } } \right )^{\Z1/3}
\ {\rm eV}
\ .
\end{equation}

Values of $(E/A)_{\rm max} \sim 10^{14} (Q/A)$ eV can be obtained for 
$f \sim 0.3$, $\eta \sim  1$ (\ie the Bohm 
limit), and $B_1 \sim 3\x{-6}$ G over a fairly  wide range of
$\tSNR$. In the examples presented here, we arbitrary set 
the parameter $f B_1 /\eta$ so that a maximum proton energy of $\sim 
10^{14}$ eV is obtained in all cases.

Unfortunately, the situation is more complicated than this since if 
cosmic rays carry enough energy to influence the shock structure, the 
shock radius and speed will, in fact, depend on the maximum energy 
cosmic rays obtain.  However, these effects are small and will not 
influence the general characteristics of the solutions we obtain.  In 
oblique shocks, there will be a far more complicated relation between 
$\eta$, $B_1$, and the shock obliquity, $\Tbn$ (\eg \EBJ 1995).  
Obliquity effects may be extremely important for modeling the radio 
emission from young SNRs, which is observed to vary considerably 
around the rim of shell-like remnants 
(e.g., Fulbright and Reynolds 1990), and may 
influence cosmic ray composition if highly oblique shocks give a 
different ratio of heavy elements to protons than do parallel ones.  
However, while we have produced results for nonlinear oblique shocks 
at nonrelativistic energies (\EBJ 1995), we are not yet able to model 
nonlinear oblique shocks to the energies required here and leave this 
to future work.

\section{NUMERICAL RESULTS}

We first produce nonlinear solutions for the shock structure including
the acceleration of protons and He$^{+2}$ to energies $\sim 10^{14}$ eV.
Both species are included self-consistently and contribute to the
smoothing of the shock. Helium is injected far upstream from the shock
at `cosmic' abundance, \iec $n_{\rm He}/n_{\rm H} =  0.1$. Once the
shock structure has been determined, we accelerate other gases and
grains as test
particles in the smooth shock, including the slowing and sputtering
of these grains
from direct collisions with the ambient gas [\ie
equation~(\ref{eq:timedirect})].  Once the grains have been accelerated, we
determine the rate at which sputtered ions are injected into the shock
and reaccelerated (as test particles)  as described above and
discussed  in
detail below.

\subsection{Non-Linear Shock Models}

We have tested several SNR models, including one with a high shock speed,
$\Vsk  =  10^4$ \kmps,
typical of a young SNR at the end of
the free expansion (or ballistic) stage (\egc \DAV 1994) (model I), 
one with an intermediate speed (\iec $\Vsk=  2000$ \kmps, model II), and one
with a slow speed typical of older, slower
remnants in the Sedov phase
($\Vsk=  400$ \kmps, model III).  
%
% zzz - 5
% 
These three models, where the parameter $f B_1/( \eta~3\mu {\rm G})$
has been set equal to 0.2 (0.34) [0.6] for model I (II) [III] to allow
acceleration of protons to $\sim 10^{14}$ eV, span a wide parameter
range and show the essential effects for most supernova explosions in
the ISM.  In addition, we illustrate the effects a low Mach number has
on the acceleration efficiency with Model IV, where $\Vsk=150$ \kmps.
In order to save computation time and improve statistics, we have used
a low cutoff energy ($\Emax \sim 10^7$ keV/nuc) and this example is not
intended to be a realistic model of SNR acceleration. 
Model IV has  $M_1=6.4$ and $r=6.6$, where $M_1$ is the far
upstream sonic Mach number.
% 
% zzz - 5

Our solutions are obtained by iteration and the technique is described
in detail in Ellison, Jones, and Reynolds (1990).  In
Figure~\ref{fig:flow}
we show the gas  flow speed in the shock frame, 
versus distance from the shock (i.e., the shock
structure or profile, $\Ux$), determined by our Monte Carlo technique
for models I, II, and III.  Notice that the distance is plotted with a
logarithmic scale for $x<-10\Lz$ and a linear scale for $x> -10\Lz$.
The shock is smoothed on the diffusion length scale $\sim \kappa/\Vsk$
of the highest energy particles in the system. Despite this extreme
smoothing, a distinct subshock persists with an abrupt transition to
the downstream state occurring in about one thermal ion gyroradius.
While the three cases shown differ in details, they are qualitatively
the same and result in similar particle spectra as discussed below.

In analyzing Figure~\ref{fig:flow}, it is essential to realize that the distance
unit $\Lz$ is proportional to $\Vsk$.  Since approximately the same maximum
energy is obtained in each case, the precursor length in real units scales 
essentially as $1/\Vsk$.

Another important point to notice in comparing the nonlinear solutions 
to the test particle one (shown as a dotted line in Figure~\ref{fig:flow}), is that 
the overall compression ratio is well above four in the  nonlinear cases. As
has been described before (\egc Ellison and Eichler 1984; Jones and Ellison
1991), in steady-state
shocks the overall shock compression will depend on the fraction of
pressure contributed by relativistic particles and on the amount of
energy flux lost at the FEB. The solutions depend on the compression
ratio in a strongly nonlinear fashion and our method determines the
overall compression as well as the shape of the flow profile
self-consistently.
The fraction of pressure that ends up in relativistic particles
depends on the shock speed as well as the Mach number 
and this contributes to the fact that Models I and II have approximately the
same
compression ratios even though they differ in Mach number.
A large compression ratio will result in flatter spectra and more efficient
acceleration.

\placefigure{fig:diffspec}

In Figure~\ref{fig:diffspec} we show differential flux spectra for models I,
II, and III.  The spectra are calculated in the shock frame, at a
position downstream from the shock, and measured in energy per
nucleon.  The light solid lines are the proton spectra and the light
dashed lines are the He$^{+2}$ spectra. The proton spectra are
normalized to one thermal proton injected far upstream per cm$^2$ per
sec and thermal helium is injected far upstream with $n_{\rm
He}/n_{\rm H} = 0.1$.  We will discuss the grain spectra (heavy solid
and dotted lines) below.

The result of shock smoothing is seen in the H$^+$ and He$^{+2}$
spectra; below $\sim A \mp c^2$ ($\sim 10^6$ keV/A) the spectra curve
slightly upward as the particles get more energetic.  This comes about
because as 
 the particles increase in energy, they develop a longer
diffusion length and `feel' a stronger compression ratio. Around $\sim
A \mp c^2$, kinematic effects cause a steepening as the particles
become relativistic, but above $\sim A
\mp c^2$, the upward curvature begins again. 
%
% zzz - 6
%
Since the highest energy particles diffuse across the full density
jump, $r$, just below the turnover caused by the FEB the spectra
develop slopes not too different from that expected from test-particle
Fermi acceleration, i.e., $dJ/dE \propto E^{-\sigma}$, where $\sigma =
(r + 2)/(r-1)$ for relativistic energies.
%
% zzz - 6

\subsection{Acceleration of Interstellar
Gases}

The fact that our calculation conserves mass, momentum, and energy
fluxes, and calculates the entire distribution function, allows a
direct measure of the absolute shock acceleration efficiency.  Once we
assume that all ion species obey equation~(\ref{eq:mfp-eqn}), the smooth shock
produces different injection and acceleration efficiencies which are
increasing function of  $A/Q$.  Unfortunately, the charge state of a
given sample of interstellar gas is not well known since we don't know
the gas temperature, and it may be
influenced by photoionization by the SN explosion UV 
flash, by X-rays from the hot, shocked downstream gas, or by 
nearby stars.  In addition, the charge state of ions can change 
due to charge stripping during acceleration.

However, the acceleration process at low energies is rapid enough that
charge stripping can be ignored in the energy range where a
substantial fraction of the $A/Q$ enhancement occurs (at least up to
$\sim 100$ MeV/nucleon if singly charged anomalous cosmic rays are
produced in a similar fashion at the solar wind termination shock,
e.g., Cummings and Stone 1996).

Therefore, if all elements start with similar charge states, the
abundance of heavy ions relative to protons will be an increasing
function of  mass.  
Because of this, we believe the increase of abundance with mass seen
for gas-phase elements in cosmic rays (Figure~\ref{fig:compmass}) is a fairly
clear signature of particle acceleration in smooth shocks (\egc
Eichler 1979; Ellison 1981).

\placefigure{fig:azflat}

To obtain 
% zzz
specific  estimates of the abundance enhancements, 
we have calculated the spectra for a number of gases,
all injected with the same number density far upstream from the shock 
and all   (except H$^{+1}$) 
with a charge of $+2$, which remains  unchanged during acceleration.
A charge state of $+2$ might occur in the cool ISM ($\sim 10^4$~K) 
subjected to UV or X-ray photoionization.
Figure~\ref{fig:azflat} shows these spectra for Models II (top panel),
III (middle panel), and IV (bottom panel) and the vertical dotted
lines show where abundance ratios relative to hydrogen are calculated.
The abundance ratios vary considerably between the three shock models.
The high Mach number shock gives large ratios and all elements are
accelerated more efficiently than protons. In the lower Mach number
shocks, however, helium can actually be underabundant relative to
hydrogen, simultaneously with the heavier elements being enhanced.
This comes about because both the pre-shock temperature and the
particle $A/Q$ influence the injection efficiency.  In high Mach
number shocks, the upstream thermal speed is a small fraction of the
shock speed for all ion species so the velocity increment gained in
the first shock crossing is approximately equal to $U_1-U_2$ for all
species.  Therefore, the efficiency is nearly monotonic in $A/Q$. In
low Mach number shocks, however, the thermal speed becomes comparable
to the shock speed and differences in the pre-shock temperature become
more important. Since we inject all species at the same temperature,
heavy ions have a smaller pre-shock speed than protons, both $A/Q$ and
the thermal speed are important and influence the acceleration in
opposite ways, and the efficiency need not be a monotonic function of
$A/Q$.  We repeat that, for computational reason, Model IV has a lower
maximum energy cutoff than the other models and is not intended to
represent a realistic SNR model. We include it to emphasize the effect
Mach number has on enhancement and to illustrate that the H/He ratio
can be greater than one over the entire energy range (bottom panel,
Figure~\ref{fig:azflat}). While the qualitative nature of the
abundance ratios shown by Model IV are largely independent of the
maximum energy, the enhancements of heavy elements are exaggerated
somewhat by the low cutoff energy which produces a larger
compression ratio than would be the case if a higher cutoff was used
(see Ellison \& Reynolds 1991 for details).
% zzz

While the ratios are energy dependent and the relative abundances of
the heavy ions continues to increase relative to hydrogen beyond 1
GeV/A (for Models II and III), in reality, charge stripping during the
longer acceleration times at higher energies will progressively reduce
the $A/Q$ values at higher energies.  We expect this effect to become
dominant around 100-1000 MeV/A, and assume that the high energy
enhancements are roughly those calculated for constant $Q=+2$ at these
energies (vertical dotted lines in the top two panels of
Figure~\ref{fig:azflat}).
% zzz
Clearly, the rules we have just stated for determining the
abundance ratios, constant $Q=+2$, ratios calculated at 100-1000
MeV/A, ignoring enhancements that occur at higher energies, are
somewhat arbitrary and other assumptions could be made; however, none
would qualitatively change the results.

\placefigure{fig:azratio}

% zzz
The abundance ratios determined from Figure~\ref{fig:azflat} are compared to
the gas-phase element cosmic ray observations
in Figure~\ref{fig:azratio}.  
The  dotted lines are from Model II ($\Vsk=2000$ \kmps) and 
the dot-dashed lines are from Model III ($\Vsk=400$ \kmps); in each
case, the upper line is calculated  at 1 GeV/A, the lower line is
calculated at 100 MeV/A, and the value for hydrogen is set to one.
The two averages from the dotted and dot-dashed lines are  shown in 
Figure~\ref{fig:compmass} with a slight renormalization.  
% zzz
Despite the uncertainties involved for the charge
state and other approximations, it's clear that the shock model does
an excellent job of reproducing the abundances of gas-phase elements
in cosmic rays. The general increase with mass is reproduced as is the
magnitude of the abundance enhancement relative to hydrogen (we note
that a similar relationship was obtained by Ellison 1981).
% zzz
Even the fact that the observed H/He ratio is actually more in cosmic
rays than in the Sun (see Paper I) can be naturally accounted for if
the shocks producing the bulk of the cosmic rays are of sufficiently
low Mach numbers, as shown by Model IV in Figure~\ref{fig:azflat}.

We also note that exceptions to the general increase of abundance with
mass may occur, 
% zzz
as with carbon, oxygen, and $^{22}$Ne, if an additional source of
material (i.e., Wolf-Rayet stars; see paper I) is present. In such a
case, the abundance will lie {\it above} our predictions. Our estimate
for the non-Wolf-Rayet abundance of cosmic ray carbon is indicated in
Figure~\ref{fig:compmass}.
% zzz

\subsection{Grain Sputtering and Abundances of Refractory Cosmic Rays}

Having determined abundances for cosmic ray gas-phase elements, we now
determine the abundance of cosmic ray refractory elements from ions
sputtered off grains.  Clearly this is a very complicated problem
since grains come in many sizes, with varied compositions, and largely
unknown structures.  Our aim here is to obtain quantitative results by
making simple, straightforward assumptions and approximations for
grain and shock properties.  This will show the plausibility that
refractory element cosmic rays originate in grains and we leave to
later work more complex models.

In our model, the same shock which accelerates interstellar gases will
simultaneously accelerate dust grains. In Figure~\ref{fig:diffspec}, the heavy
solid lines are test-particle `grain' spectra for grains with our
standard parameters, i.e., $a=10^{-7}$ m, $\phi = 10$ V, and $\mu = 56$
(yielding $A/Q \simeq 8\x{7}$). The grains have been accelerated in the
smooth shocks shown in Figure~\ref{fig:flow}. They  have undergone losses from
collisions with the gas and
the high energy 
turnover reflects the situation where the loss time approximately equals
the acceleration time. This should be compared to
equation~(\ref{eq:energymax}). Note that we take $B_1=3\x{-6}$ G and
$\dengas=1$ \pcc \ here and in all of the examples below.
The test-particle grains were
injected with $\denGr = \denH$ and must be scaled by the actual ambient
thermal grain density to obtain the absolute normalization (i.e.,
$\denGr/\denH \sim 3\x{-14}$). 

The enhancement effect for large $A/Q$ particles is clearly seen in the
grain spectra. While the grains were 
injected with the same number density as
protons, they obtained 
a much flatter spectrum at low energies resulting in a
substantial enhancement before their spectra cut off. As they accelerate,
the grains  sputter and
ions sputtered off in the upstream region will be further accelerated
upon convecting into the shock. We model this by including in the \mc
simulation a direct determination of $I(v)$ during grain acceleration
and then perform a separate run using $I(v)$ as the injection rate of
grain products. In reality, this reacceleration would occur simultaneously
with the grain and gas acceleration. 

Since the Monte Carlo code follows individual particles, we know the time
spent in the upstream region, the position
of the particle, its speed,  and the  
background density at that position. We
find $I(v)dv$ by multiplying the
sputtering rate, $0.01\dengas(x) a^2 v_{\rm G}$, by the time and
summing the number of sputtered atoms produced in each velocity bin.
We neglect all ions sputtered downstream from the shock
since a sputtered ion has an $A/Q$ ratio many orders
of magnitude smaller than its parent grain, which means it is many
more mean free paths downstream from the shock than the parent grain.
We sum over the
entire upstream region assuming, as before,  that all
upstream sputtered ions are injected into the shock without losses.
In this way we
obtain $I(v)$, the number of injected
ions per unit area per unit time with speeds
between $v$ and $v+dv$.

When $I(v)$ is injected and accelerated, we obtain the results shown
with heavy dotted lines in Figure~\ref{fig:diffspec}. Here we have assumed that
the sputtered ions have $A= 56$ and $Q= 2$, but show below the effect
of varying this charge.  Again, the sputtered ion spectra must be
scaled down by the actual ambient grain density since we have injected
the grains with $\denGr= \denH$.  The most striking feature in
Figure~\ref{fig:diffspec} is that the sputtered product flux lies many orders
of magnitude above the grain flux. This comes about mainly because
each grain contains $\sim 10^9$ iron atoms, so that even a small
fraction ($\sim 10^{-3}$) of sputtered ions will make up a large
flux of accelerated ions. An additional increase comes in because the
sputtered products with $A/Q=56/2$ are accelerated more efficiently
than the protons or alphas.  Charge stripping during acceleration will
lower this enhancement somewhat.

\placefigure{fig:injrates}

The injection rates, $I(v)$, for the three examples of
Figure~\ref{fig:diffspec}, still normalized to $\denGr = \denH$, are shown 
in 
Figure~\ref{fig:injrates}.
As described in Section~2.4.1, $I(v)$ peaks strongly just below the
grain cutoff energy and we use this fact to approximate the injection
of grain erosion products as a $\delta$-function at the weighted mean
energy of $I(v)$.  The actual sputtered ion injection energies per
nucleon, $E_{\rm inj}/A$ and rates, $f_{\rm inj}$, used in
Figure~\ref{fig:diffspec} are labeled in Figure~\ref{fig:injrates}.

The acceleration of grains and sputtered products is an extremely
complicated process with a number of factors influencing the final
cosmic ray abundance. These factors include: the size, charge, and
mean molecular weight of the grain, the background gas density, the
collision and sputtering rate, any losses sputtered ions experience
before being further accelerated, and the charge state of the
sputtered ion including charge stripping before and during
acceleration. In addition, shock properties, such as the Mach number,
shock age, geometry, ambient magnetic field strength, and maximum
proton energy obtained, will modify the resultant cosmic ray
abundance.  Since many of these factors are poorly known, it is
impossible to precisely predict the refractory cosmic ray abundance.
However, we can investigate some of these factors to see how robust
the process is.

\placefigure{fig:diffspecbig}
In Figure~\ref{fig:diffspecbig} 
we show results for grains of various sizes. We have
used Models I ($\Vsk=10^4$ \kmps) and III ($\Vsk=400$ \kmps) and the lower
eight panels show the low energy portion of the proton
spectra (light lines) with `grain' spectra (heavy lines) of 
sizes, $0.001  < a < 1 \mu$m, all injected with the same far upstream
number density as protons and all including losses by collisions with
the background gas. We have kept all
other parameters constant and equal to those of Figure~\ref{fig:diffspec},
i.e., $\mu =  56$, $\phi= 10$ V,  $B_1= 3\x{-6}$ G, and $\dengas = 
1$ \pcc.
Several important effects are illustrated by these plots. 
The top two panels, labeled $A/Q= 2$, show a comparison between protons
and helium (we repeat that in these plots all species are injected
with the same far upstream number density) 
and there is barely any enhancement effect for helium over protons.
However, as $a$ and
$A/Q$ increase, the slope of the distribution flattens
and the enhancement becomes quite large. 
For each horizontal pair of panels, the $A/Q$ values shown are
obtained from equation~(\ref{eq:AoverQ}) using the grain size labeled in each
panel. 
It is also clear from this figure that as $A/Q$ is
increased, the enhancement reaches a maximum (at any given
energy per nucleon)
and then falls off for larger $A/Q$. At some point
(e.g., $a\sim 1\mu$m for Model I and $a>1\mu$m for III),
the grain size becomes
large enough that the grains essentially only cross the shock once 
before
being lost downstream.
For some grain sizes, the
enhancement over protons is more than a factor of 100 at a few MeV/A
and this will translate into the
efficient production and  acceleration of sputtered ions. 
The important point is that 
significant enhancement of grains occurs for sizes 
spanning at least three orders of magnitude in radius (i.e., more than six
orders of $A/Q$) for very different shock speeds  ranging from
$\Vsk=400$ to $10^4$ \kmps \  indicating that
the effect is quite robust. It is clear, however, that very small or very
large grains will not be enhanced by this process.
The results for Model II lie between those shown.

Another aspect of the grain acceleration evident in
Figure~\ref{fig:diffspecbig} is that as $A/Q$ is increased, the downstream
quasi-thermal peak shifts to higher energy per nucleon (the
quasi-thermal peak is made up of ions or grains that have crossed the
shock only once).  This effect also comes about because of the smooth
shock. If the shock were discontinuous, all spectra would show the
thermal peak at approximately the same energy per nucleon, the only
difference coming from differences in upstream thermal speeds which
will be quite small for high Mach numbers. However, in the smooth
shock, larger $A/Q$ particles get a larger velocity kick on their
first crossing of the shock. For $\Vsk= 10^{4}$ \kmps\ (Model I), the
velocity kick received for $A/Q \sim 10^8$ grains is approximately
five times that for $A/Q= 1$.

We now make specific assumptions for the shock and grain properties in
order to obtain a direct estimate of the cosmic ray abundance of iron.
Our predictions, of course, will  depend on the particular
assumptions we make.  We assume that {\it all}  iron which
ends up as cosmic rays originates in grains. Using our $400$
\kmps \ shock Model III, 
we inject and accelerate grains with $a= 0.1\mu$m, $\phi= 10$ V, $\mu
= 56$, assuming $B_1 = 3\x{-6}$ G, and $\dengas= 1$ \pcc.  We next
assume that all sputtered iron ions are swept back into the shock
without significant energy losses. Furthermore, we assume that in the
100--1000 years or so they spend convecting back to the shock [i.e.,
equation~(\ref{eq:tad})], they become fully stripped as seems likely. The
Fe$^{+26}$ ions are then re-accelerated in the same shock which
accelerated the grains.

\placefigure{fig:ironspec}

\placefigure{fig:crdata}

Our results are shown in
Figure~\ref{fig:ironspec} and compared to the proton spectrum (light and
heavy solid lines).  The accelerated grains are shown with a dashed
line, and sputtered Fe ions are shown with dotted lines.  The
test-particle grains are injected far upstream with solar (i.e.,
``cosmic'') abundance, that is, the total number of Fe {\it atoms}
in the grains is $\sim 3.1\x{-5}$ times the number  of H atoms.
This means that the ratio of far upstream number densities is
$\denGrFe/\denH\sim 3\x{-14}$ since, for $a=0.1 \mu$m, there are $\sim
10^9$ Fe atoms per grain.  In order to indicate the difference the
sputtered ion
charge state makes, we show Fe spectra for Fe$^{+2}$ (i.e., without
stripping), as well as Fe$^{+26}$.
In the actual case, there will be a mixture of charge states since
sputtered ions formed close to the shock will be convected back before
much charge stripping can occur.  These ions will be accelerated more
efficiently than the Fe$^{+26}$ ions, but fully stripped ions obtain a
higher maximum energy per nucleon in the finite size shock.  On the
other hand, sputtered ions will experience some ionization energy
losses before encountering the shock which will cause them to be
accelerated less efficiently. In want of a more complete model, we
compare the Fe$^{+26}$ spectrum with no upstream losses to the
observations 
in Figure~\ref{fig:crdata} below.

In any event, it is clear from this figure that a large
enhancement of iron over protons occurs. Between $\sim 10$ and 100
GeV/A, the sputtered Fe$^{+2}$ ions
stand about a factor of $10^3$ above their solar abundance relative to
hydrogen and the Fe$^{+26}$ stands about a factor of  $20$  above
(compare the light solid proton line, which has been divided by the cosmic
abundance of Fe, to the heavy dotted lines).
In the next section we compare these predictions, both normalization and
spectral shape, to observed cosmic ray spectra.

\subsection{Spectral Comparisons}

In Figure~\ref{fig:crdata} we compare our proton, helium, and Fe$^{+26}$
spectra to cosmic ray observations.  The data in Figure~\ref{fig:crdata} is
adapted from the compilation of \cosray observations presented by
Shibata (1995). Both data and model spectra have been multiplied by
$(E/A)^{2.5}$ to flatten the steep spectra.  

In addition, since we compare with observed spectra, we must correct
our calculated source spectra for rigidity dependent escape from the
galaxy
% zzz
and reduction due to nuclear destruction. 
% zzz
Evidence, including the comparison of secondary to primary element
spectra, indicate that \crs escape from the galaxy at a rate
proportional to some power of the rigidity, \ie $R^{-\delta}$ at
relativistic energies (\egc Protheroe, Ormes, \& Comstock 1981;
Engelmann \etal 1990; Shibata 1995).  While the actual escape may be
more complicated than this and the exponent may vary somewhat with
different galactic propagation models and/or energy, this form is
sufficient for our purposes considering other uncertainties in our
model.  With this assumption, the ratio of the observed flux to the
source flux for a species $\alpha$ is
\begin{equation}
{\Phialpobs \over \Phialps} \propto R^{-\delta} 
\ .
\end{equation}
If we consider only relativistic particles and compare fluxes of
species $\alpha$ to protons we have,
\begin{equation}
{
\left ( {\Phialpobs \over \Phialps} \right ) 
\!\! \biggm \slash \!\!
\left ( {\Phipobs \over \Phips} \right )
} = 
{
\left ( { \Phialpobs \over \Phipobs } \right ) 
\!\! \biggm \slash \!\!
\left ( { \Phialps \over \Phips } \right )
} =
\left ( {R_{\alpha}\over R_{\rm p}} \right )^{\Z-\delta}
=
\left ( {A_{\alpha} \over Q_{\alpha}} \right )^{\Z-\delta} 
\quad {\rm at \ equal} \ {E \over A} 
\ ,
\end{equation}
and
\begin{equation}
= 
\left ( Q_{\alpha} \right )^{\delta} 
\quad {\rm at \ equal} \ E \ {\rm per \ nucleus} 
\ .
\end{equation}
Thus, at fully relativistic energies,
identical source spectra in energy per nucleon will show He$^{+2}$
lower by a factor $2^{-\delta}$ after energy dependent escape from the
galaxy.
If additionally, protons are 10 times more numerous in the source gas
than helium nuclei, the observed helium flux will be $2^{-\delta}/10$
as intense as the proton flux assuming both are accelerated with the
same efficiency.

% zzz
We have corrected the helium and iron fluxes for nuclear destruction
using for the surviving fraction, $1/[1+ (\lambda_e/\lambda_d)]$,
where $\lambda_e$ is the rigidity dependent escape length and the
nuclear destruction length, $\lambda_d$, is 2.7 g cm$^{-2}$ for iron
and $\sim 35$ g cm$^{-2}$ for elemental helium.
% zzz

For the comparison in Figure~\ref{fig:crdata}, we have used Model III
($\Vsk= 400$ \kmps) and our results for protons are shown as a
solid line and those for He$^{+2}$ with a dashed line.  The model source
spectra  have been  multiplied by $R^{-0.65}$
% zzz
and helium has been corrected for nuclear destruction, although this is
a small effect for helium.
The rigidity dependence we have chosen, $\delta=0.65,$ is close to the
best value, $\delta=0.6$, adopted by Engelmann \etal (1990) and
Shibata (1995). If reacceleration during propagation is important,
$\delta$ could go down to 0.3 (e.g.,  Berezinskii \etal 1990).

Several features are evident in Figure~\ref{fig:crdata}. First of all, solar
modulation is not included in our model and its effect shows in the
data which fall off somewhat below the kinematic turnover near 1
GeV/nucleon. Second, our model matches the gross features of the
spectra extremely well above the energy where modulation is important
and below where our model spectra fall off from the effects of our
adopted finite size SNR shock. We must caution that the fit to the
slope depends strongly on the value for $\delta$ we choose and should
not be over interpreted. Cosmic rays undoubtedly come from a number of
supernovae of varying sizes, Mach numbers, etc., and a model such as
ours of a single shock is not intended to model the observations in
complete detail.  Our Models I and II, with larger compression ratios,
produce even flatter spectra than Model III and either would not fit
the spectral observations or would fit only with a larger value of
$\delta$. In addition, they would predict H/He ratios that are too
low.  Having said this, we wish to emphasize while we have adjusted
the overall normalization of the {\it proton} spectrum to match the
observations, there is {\it no adjustment of the relative
normalization} between protons and helium. Our 400 \kmps \ shock model,
using the cosmic abundance of helium, reproduces the observed relative
fluxes fairly accurately. The fact that our model yields a H/He ratio
somewhat lower than observed, may imply that lower speed shocks are
important (see Figures~\ref{fig:azflat} and \ref{fig:azratio}). 
Lastly, the proton and
helium spectra from the nonlinear model show a slight upward curvature
indicative of nonlinear shock acceleration.

The 
% zzz
heavy 
% zzz
dotted line in Figure~\ref{fig:crdata} shows the Fe$^{+26}$ curve from
Figure~\ref{fig:ironspec} multiplied by $(E/A)^{2.5} \times R^{-0.65}$
% zzz
and corrected for nuclear destruction.  
% zzz
Once these
corrections are made,
% zzz
the
normalization relative to protons is exactly that shown in
Figure~\ref{fig:ironspec}, \iec the iron is injected out of a medium with
solar abundance
$\denFe = 3.1\x{-5} \denH$ and the model sets the abundance
relative to hydrogen.  Clearly, the excellent match to the
observed normalization is somewhat fortuitous considering the many
uncertainties in the model. However,  we have chosen
parameters we feel to be realistic and demonstrated that the grain model is
quite robust; we do not expect that this result
will fundamentally change as refinements are made.  
% zzz
The light dotted line shows the Fe$^{+26}$ spectrum without the 
nuclear destruction correction  which tends to remove the otherwise 
distinctive
curvature expected from nonlinear shock acceleration.
Without this correction, the non-power law nature of the spectrum 
is more pronounced and, in fact, we predict that this
curvature is real and sufficiently accurate cosmic ray observations
may reveal it.
% zzz

% zzz
Our prediction for the source cosmic ray iron abundance is included in
Figure~\ref{fig:compmass}.  To give some indication of the errors intrinsic to
our calculation, we show two horizontal lines on the right side of the
plot.  The lower line is the ratio, Fe$^{+26}$/H$^+$, taken at 100
GeV/A in Figure~\ref{fig:ironspec}, and the upper line is taken at 10 GeV/A.
The match is excellent. The same enhancements should apply roughly for
the other refractory elements in Figure~\ref{fig:compmass} since, in the
crucial early acceleration phase, they are all accelerated, not as
individual ions, but as constituents of the same grains.
% zzz

The obvious failure of our model to produce spectra above $\sim 10^6$
GeV is one that all models using single, isolated SNRs have. 
% zzz
While parameters can be chosen to extend the maximum energy up above
$10^{16}$ eV, it has been known since Lagage and Cesarsky (1983) that
{\it standard}
% zzz
parameters for shock acceleration and SNRs yield a maximum energy
below the observed knee in the cosmic ray spectrum and far below the
highest energy cosmic rays. More elaborate models involving explosions
into the pre-supernova stellar wind may be able to account for cosmic
rays up to and beyond the knee (see V\"olk and Biermann 1988; Biermann
1993).

\section{SUMMARY AND DISCUSSION}

\subsection{GCR Source Composition; Volatility Versus FIP}

We have presented a one-site, one-step model of galactic cosmic 
ray (GCR) origin and acceleration that produces cosmic ray 
source spectra with slopes and relative abundances which match 
observations.
In this scenario, GCRs come from interstellar or circumstellar 
gas and dust grains accelerated simultaneously by SNR shocks.
Our model combines an acceleration of the 
gas-phase, volatile elements with an $A/Q$ dependent enhancement, and 
a preferential acceleration of the grains, and hence of the sputtered, 
refractory elements initially locked in them.

In addition to accounting for the general excess of the standard 
low first ionization potential (FIP) 
refractory elements, this scenario can account naturally for 
a number of the long-standing puzzles in the GCR source composition 
(see Paper~I):
(i) The low H, He, and N abundances, relative to heavier 
elements  ($A/Q$ effect);
(ii) The currently assessed low Na/Mg, Ge/Fe, Pb/Pt, and the 
high P/S ratios (grain acceleration effect);
(iii) The weak mass-dependence of the refractory element enhancements, 
as opposed to the strong mass-dependence of the enhancements for the 
volatiles (since the refractory elements are first accelerated, 
not as individual ions, but as constituents of grains);
(iv) The apparent general overabundance of ultra-heavy elements beyond 
$Z \sim 60$ or even $\sim 40$, if it is confirmed ($A/Q$ effect); and
(v) The presence of $^{22}$Ne, C, and O enhancements in GCRs, which 
comes naturally from the contribution of the most massive SNae, which 
accelerate their own $^{22}$Ne--\cto--\os enriched pre-SN Wolf-Rayet 
wind material.

Our predictions are summarized in Figure~\ref{fig:compmass}.  
% zzz
The dotted,
dot-dashed, and solid  curves
 show our predictions 
% zzz
for highly volatile cosmic ray abundances as a function of mass (in a
specific ionization model) for shock velocities of 2000, 400, and
150 \kmps, respectively. The somewhat high H/He ratio observed
suggests a significant contribution of lower Mach number shocks. The
horizontal solid lines give our estimates for refractory element (\ie
iron) GCRS abundance, which is more or less independent of mass.  We
interpret the elements with intermediate volatility as coming from a
mixture of gas and dust in the ISM.

Specifically, we predict that:
(i)  All gas-phase elements accelerated out of 
ISM or circumstellar matter with normal composition will have 
abundances which lie 
% zzz
in the range given by the dotted and dot-dashed lines
% zzz
shown in Figure~\ref{fig:compmass}.
Based on the observed \nett excess, we expect an additional 
source of carbon and oxygen from the acceleration of 
\cto--\oso--enriched Wolf-Rayet wind material affected by He-burning 
nucleosynthesis, causing these elements to lie {\it above} the line by
the amount contributed by this additional source.  Our estimate for
the non-Wolf-Rayet carbon contribution is labeled in Figure~\ref{fig:compmass};
it could represent an overestimate if a significant fraction of carbon
is not in the gas phase (as we assumed), but locked in grains in the
C--rich Wolf-Rayet wind material (C/O $>1$), and hence preferentially
accelerated.  (ii) Refractory elements such as iron, magnesium,
silicon, etc., which are essentially locked in dust grains in the ISM,
will be accelerated preferentially relative to the gas-phase elements,
and will have GCRS abundances determined by grain properties and
gas-grain interactions, such as mean grain size, sputtering rates,
etc.  However, we have shown that the preferential acceleration of
heavy grains is not strongly dependent on these properties.  So, all
refractories should end up with abundances independent of condensation
temperature and mass, and not too far from our prediction for iron.
(iii) No significant amount of SN ejecta material is being
accelerated, in view of the minor role played by the short-lived
reverse shock, and of the low probability that fast blobs of ejecta
material overtake the forward shock.  (iv) Cosmic ray spectra (at
energies above those where solar modulation is important) will not be
strict power laws, but will show a concave upward curvature
distinctive of nonlinear shock acceleration.  
(v)
The shock acceleration of interstellar grains will produce grain
speeds relative to the background plasma considerably greater ($\betaG
\sim 0.01$) than is generally assumed.  While only a small fraction of
the total number of grains will obtain these maximum speeds, the bulk
of the grains will be shock heated to $\sim$keV/A energies (see
Figure~\ref{fig:diffspec}). This may result in important modifications
to the X-ray modeling of the shock wave environment, and may explain
recent observations of a broad $^{26}$Al $\gamma$-ray line. The width
of the 1.809-MeV $\gamma$-ray emission line seen by GRIS (Naya \etal
1996), implies that $^{26}$Al decay occurs at speeds $>450$ \kmps, or
at energies per nucleon of $\sim 1$ keV/A. If the $^{26}$Al is locked in
grains with large rigidities, shocks with speeds somewhere between our
Models II and III (see Figure~\ref{fig:diffspec}), will produce 
grain distributions consistent with these results.

For many years, the GCR source composition has been interpreted 
in terms of a FIP fractionation, similar 
to that extensively observed in the solar environment, but usually 
without considering the possibility of an $A/Q$ dependence of the 
acceleration efficiency (Meyer 1985; see, however, Silberberg and Tsao 
1990).
One may wonder whether a FIP fractionation, combined with an $A/Q$ 
dependent bias reflecting acceleration conditions, could account 
for the composition data as convincingly as our volatility model.
This hypothesis would equally well explain the strong mass 
dependence of the high-FIP (volatile) element enhancements,  
but we do not see how it could simultaneously account for the 
lack of a comparable, strong mass-dependence of the low-FIP 
(refractory) element enhancements.
Indeed, 
in any FIP scenario the particles have to be accelerated out 
of an already FIP-biased gas, resulting from a prior ion-neutral 
fractionation; in this gas, all elements, whatever their FIP, have to 
be at least singly ionized to get accelerated, so that low- and 
high-FIP elements no longer behave differently (cf., e.g., solar 
energetic particles accelerated out of FIP-biased 10$^{6}$~K coronal 
gas; \eg Meyer 1985, 1993).
The combined FIP and $A/Q$ hypothesis cannot account for the low
Na/Mg, Ge/Fe, Pb/Pt, and high P/S ratios since two elements of
comparable FIP and mass are being compared in each of these ratios.
Furthermore, FIP scenarios require a two-stage acceleration mechanism,
in two different sites (acceleration to MeV energies in later-type
star environments; combined with acceleration to GeV and TeV energies
by SN shock waves), and the Wolf-Rayet star source for the
$^{22}$Ne-C-O excess has to be treated as a totally separate component
(Meyer 1985).  So, in our current view, the great 
similarity\footnote{The similarity is actually not complete
since Na and P seem to be less abundant in the GCRs than in solar
energetic particles (Garrard \& Stone 1993; Reames 1995; paper I).}
between
the GCR source composition (volatility-biased, due to preferential
acceleration of grain material), and the solar coronal, solar wind,
and solar energetic particle composition (FIP-biased, due to an
ion-neutral fractionation in the $\sim $ 10$^{4}$~K solar
chromosphere) is purely coincidental!

\subsection{Other Non-Linear Shock Acceleration Models}

While we believe ours is the first attempt to give a detailed 
description of grain and gas acceleration in nonlinear shocks, a great 
deal of work on nonlinear acceleration of protons (and some including 
other atomic species) has been done 
(for reviews, see Blandford and Eichler 1987; Berezhko and
Krymsky 1988; Jones and Ellison 1991).  The 
work that is closest to ours has been presented in a series of papers 
by Berezhko and co-workers (Berezhko, Yelshin, and Ksenofontov 1994; 
Berezhko, Ksenofontov, and Yelshin 1995; Berezhko, Elshin, 
and Ksenofontov 1996),
but see Kang, Jones, \& Ryu (1992), and Kang \& Jones (1995, 1996)
for similar nonlinear shock work.  Berezhko et al.
couple the diffusive transport equation describing the cosmic ray 
distribution function to the background fluid described by gas dynamic 
equations with the cosmic ray pressure added in.  This is much the 
same as done in \DAV (1993) and a number of previous investigations
(see above reviews).
However, unlike 
most previous analytic studies, Berezhko \etal have been able to 
develop techniques which allow them to use a strongly energy dependent 
diffusion coefficient.  As done here, they assume that the scattering 
mean free path is proportional to the gyroradius, but set $\kappa= \rg 
c/3$ for simplicity, neglecting the change in particle speed at 
nonrelativistic energies.  
They 
determine the fraction of the total ejecta energy going into 
cosmic rays as a function of time, and calculate the overall spectrum 
of cosmic rays during the evolution of the remnant, including 
adiabatic losses as the high density region near the shock expands to 
the ISM value. The result is a spectrum which is quite similar to 
a power law in momentum over the entire range from thermal to $\sim 
10^{15}$ eV, and with about 20\% of the ejecta energy going into 
cosmic rays.

One major difference between their work and ours is that they have a 
time-dependent model and are, therefore, able to follow the evolution 
of the shock and the SNR dynamics, whereas we are restricted to a 
steady-state, and do not calculate the overall remnant dynamics.  
However,  they 
conclude
that geometric factors, \iec particle escape from the finite 
sized shock, limit the maximum energy obtained in the shock.  
This fact makes our steady-state model more applicable than would be 
the case if remnant age limited acceleration.

Another important difference between our model and those 
based on the diffusion equation is that the 
diffusion approximation (\iec the requirement that particle speeds be 
large compared to flow speeds) limits the ability to treat particle 
injection.  For example, Berezhko \etal 
must assume that some 
small fraction, $\epsilon$, of the incoming gas is 
transferred to cosmic rays, and the overall efficiency with 
which cosmic rays are produced, $E_{\rm cr}/E_{\rm SN}$, depends 
critically on $\epsilon$.  In Berezhko, Ksenofontov, and Yelshin 
(1995), $E_{\rm cr}/E_{\rm SN}$ is shown to vary from about 0.2 for 
$\epsilon \simeq 10^{-4}$ to about 0.8 for $\epsilon \simeq 10^{-3}$ 
for a shock with an initial Mach number of about 30.
  
In contrast, our \mc description is not restricted to superthermal 
particles, and once we assume that all particles obey 
equation~(\ref{eq:mfp-eqn}), injection is treated self-consistently.  There 
may be questions concerning the appropriateness of 
equation~(\ref{eq:mfp-eqn}), but once such a scattering description 
has been chosen, both the injection rate and the injection 
momentum are fully determined by the solution, for all species, 
without any additional 
parameters such as $\epsilon$.
In contrast, Berezhko, Elshin, and Ksenofontov 
(1996) must make two additional assumptions
% zzz
to treat species other than protons.
% zzz
The first is that all species obtain the same thermal velocity 
distribution {\it behind} the shock, and the second is an ad hoc 
enhancement factor, $e(A/Q)= R^\beta$, added to the injection rate,
where $\beta$ is a free parameter chosen  to match cosmic 
ray abundance observations. 

\subsection{Future Work}

The current study provides a framework that seems qualitatively
capable of accounting for all the features of the observed composition
within a single acceleration context.  Much work is required to
substantiate it, and to ensure that it can apply in the real supernova
(SN) shock wave environment.  In particular, a number of points will
have to be investigated for the various stellar masses contributing
SNae, according to their weight in the initial mass function and to
their efficiency in accelerating particles.  These include: (i) The
nature of the external material accelerated by the shock.  This is
presumably local ISM for the lower mass SNae, and pre-SN stellar wind
for the more massive ones (including the Wolf-Rayet stars responsible
for the $^{22}$Ne-C-O enhancements); (ii) The ionization states
present in this external material.  It could be $10^{6}$~K material,
in which refractory grain cores would not have been evaporated; it
cannot be collisionally ionized $10^{4}$~K material, in which Ne and
He would be mainly neutral, but it could be $10^{4}$~K photoionized
material, \egc by the UV burst associated with the SN explosion, or by
an X-ray precursor associated with the shock wave.  The more precise
relationship between $A/Q$ and $A$ for the volatile elements should
be, as much as possible, investigated in the various hypothesis; (iii)
The origin of the grains, which could be those formed recently in the
pre-SN stellar wind for the massive SNae, or old ISM grains for the
lower mass ones.  The grain composition may be different for grains
newly formed in stellar winds, in which \Tc could be the essential
parameter, and for old ISM grains, where slow chemical reprocessing
should be important as well (\egc Jones \etal 1994; Draine 1995); in
the old ISM, phosphorus, in particular, seems less locked in grains
than expected based on its rather high \Tco, which is not consistent
with its presumably high GCRS abundance; this might represent a hint
that circumstellar dust is important in the GCR sources (\egc Savage
and Sembach 1996; Paper I); (iv) The expected excesses of $^{22}$Ne,
C, and O associated with the contribution of the most massive, WC and
WO Wolf-Rayet star SNae should be more precisely evaluated, as well as
the, certainly much smaller, excess of nitrogen associated with WN
Wolf-Rayet stars; note that the carbon excess (Figure~\ref{fig:compmass}) might
be due, in addition to Wolf-Rayet star nucleosynthesis, to a
significant fraction of carbon being condensed into solid form in
C-rich Wolf-Rayet atmospheres (C/O $>$ 1), where {\it not all\/} C is
prevented from condensing by CO formation (\eg van der Hucht and
Williams 1995).  For more details, see Paper~I; (v) The determination
of the relative contribution of SNRs with various shock strengths in
order to simultaneously account for both the systematic increase of
the enhancement with mass among the volatile elements and the somewhat
high H/He ratio.

\section{CONCLUSIONS}

A re-analysis of the observed galactic cosmic ray chemical composition 
(especially Na, P, Ge, and Pb; Paper I)  strongly suggests
that cosmic ray source material  consists mainly 
of two components both originating in the interstellar medium (and/or 
circumstellar material):
volatile elements from  the gas-phase, and refractory elements
from dust grains.
Relative to solar  abundances, the abundances of 
the volatile elements are found to be a strongly 
increasing function of mass with the sole exception of hydrogen.
In contrast, 
the abundances of the refractory elements, known to be
locked in grains in 
the ISM, are systematically higher, but with  little or no mass 
dependence, allowing a clear separation of these components in the 
data.
The elements which are  likely to be {\it partly\/} locked in 
grains in the ISM show up midway between these two groups.
This evidence points to separate injection and/or acceleration
processes depending on whether an element is mainly in the gas-phase
or in dust in the ISM.

We have shown here that standard nonlinear shock acceleration theory 
can account for  these composition features.
We assume that supernova remnant blast waves moving through the 
undisturbed ISM  or circumstellar matter pick up and accelerate 
gas  ions and dust  grains simultaneously.
The gaseous ions are accelerated directly to cosmic ray 
energies in the smoothed 
shock, which produces an enhancement of high 
mass/charge (\ie $A/Q$) elements.
Since heavier elements always tend to have higher $A/Q$ ratios, 
the mass dependence seen in the volatile  cosmic ray abundances 
is a clear signature of acceleration by  smoothed 
shocks. This accounts, in particular, for the low cosmic 
ray hydrogen and helium abundances relative to heavier volatile 
elements; the somewhat high cosmic ray H/He ratio suggests a
significant contribution of low Mach number shocks in the acceleration
process, consistent with the estimated source spectral shapes.
If we assume that the  weakly charged,  massive grains act in the 
ambient magnetic fields exactly as protons of the same rigidity, these 
grains  will be very efficiently accelerated by the same shocks, 
although  up to far lower energy per nucleon ($\sim 
$~100~keV/nucleon) than the gas  
ions, due to friction and to the limited age and size of the SNR.
By including a simple model of the sputtering of these grains upstream 
from the shock, and the acceleration of the 
sputtered ions to  GeV and TeV cosmic ray energies, we 
are able to calculate the relative abundances of refractory
elements to volatile ones.  
In view of the crucial role of the injection stage in shaping 
the composition, this injection of refractory elements as constituents 
of entire grains results in a lack of a significant mass dependence of 
the refractory element abundances, as observed.

So, the elements originating in gas and dust are both accelerated by
the same shocks, but have different abundances in cosmic rays because
they have different injection routes; gas-phase elements are directly
picked out of the thermal plasma and accelerated, while grain material
is first accelerated as entire grains, the grains are then subjected
to sputtering, and finally the sputtered ions are further accelerated,
in the same shock, to cosmic ray energies.
The relative abundances of refractory  to gas-phase elements 
is determined by  grain physics  such as 
the sputtering rate, average grain size, etc., and is found to be 
consistent with observations.

It is important to note that, if our model is correct, the long 
standing belief that  cosmic rays accelerated to GeV and TeV energies by 
SNR shock waves were first injected to MeV
energies with appropriate coronal gas composition
by later-type stars, is rejected.
This belief was based almost solely on the fact that solar energetic 
particle and  galactic cosmic ray abundances show a similar 
correlation  with first ionization potential (FIP).  This 
similarity is now assumed to be coincidental, stemming from the fact 
that there is a strong correlation for most elements between 
volatility and FIP.
We also note that our one-step, one-site model is much simpler than 
any scenario based on stellar  injection processes, which 
involve  two acceleration stages in  two unrelated sites.
Our process also 
accounts naturally for the relative deficiency of the two 
most abundant species, hydrogen and helium,
and can  account,  without  recourse to a 
different type of source, for the \netto--C--O excess observed in 
galactic cosmic rays; this excess comes from 
the most massive star SN shocks, which are 
bound to accelerate their own \netto--\cto--\oso--enriched pre-SN 
Wolf-Rayet star wind material.

The model we have presented is far from complete.  We have used a 
plane, steady-state model to mimic a curved, evolving supernova 
remnant.  Further, grain properties are sufficiently unknown 
that fairly large uncertainties exist for typical grain sizes and 
sputtering rates, and we have made no attempt to model SNRs  
developing in interstellar or circumstellar media  with different 
parameters.  We also must assume that grains interact with the 
background magnetic field nearly elastically, as is believed to be the 
case for  gas ions.  All of these and other uncertainties need 
to be investigated.  However, we believe the  striking ordering 
of the cosmic ray  composition data  in terms of refractory 
and volatile elements (Figure~\ref{fig:compmass})
is a compelling reason to require that 
interstellar grains be accelerated by shock waves.

It is also clear that we fail to explain cosmic rays with energies 
above the knee at $\sim10^{15}$~eV, and that our spectra are 
somewhat flatter than required by some propagation models.  These 
problems, however, are common to any model based on an isolated 
supernova exploding in the ISM (e.g., Berezhko, Elshin, and 
Ksenofontov 1996), and have nothing to do with the acceleration of 
grains.

\acknowledgements

D. Ellison and L. Drury wish 
to acknowledge the hospitality of the Service d'Astro\-phy\-sique, 
Centre d'Etudes de  
Saclay where much of this work was carried out.  L. Drury's 
visit was supported by the Commission of the European Communities 
under contract ERBCHRXCT940604, and D. Ellison was supported, in part, 
by the NASA Space Physics Theory Program, 
the Service d'Astro\-phy\-sique, the Observatoire de Paris-Meudon, and
CNET/CETP (Issy-les-Moulineaux).  The authors also thank T. 
Shibata for kindly furnishing recent cosmic ray data, and K. 
Borkowski, T. Gaisser,
S. Reynolds,  and D. Reames for helpful discussions.

\def\itt{\rm}
\def\bff{\rm}
\def\aa#1#2#3{ 19#1, {\itt A.A.,} {\bff #2}, #3} 
\def\aasup#1#2#3{ 19#1, {\itt A.A. Suppl.,} {\bff #2}, #3} 
\def\aj#1#2#3{ 19#1, {\itt A.J.,} {\bff #2}, #3}
\def\anngeophys#1#2#3{ 19#1, {\itt Ann. Geophys.,} {\bff #2}, #3} 
\def\annrev#1#2#3{ 19#1, {\itt Ann. Rev. Astr. Ap.,} {\bff #2}, #3}
\def\apj#1#2#3{ 19#1, {\itt Ap.J.,} {\bff #2}, #3} 
\def\apjlet#1#2#3{ 19#1, {\itt Ap.J.(Letts),} {\bff  #2}, #3} 
\def\apjpress{{\itt Ap. J.,} in press}
\def\apjs#1#2#3{ 19#1, {\itt Ap.J.Suppl.,} {\bff #2}, #3} 
\def\app#1#2#3{ 19#1, {\itt Astroparticle Phys.,} {\bff #2}, #3} 
\def\asr#1#2#3{ 19#1, {\itt Adv. Space Res.,} {\bff #2}, #3}
\def\araa#1#2#3{ 19#1, {\itt Ann. Rev. Astr. Astrophys.,} {\bff #2}, 
   #3}
\def\ass#1#2#3{ 19#1, {\itt Astr. Sp. Sci.,} {\bff #2}, #3}
\def\eos#1#2#3{ 19#1, {\itt EOS,} {\bff #2}, #3}
\def\icrcplovdiv#1#2{ 1977, in {\itt Proc. 15th ICRC(Plovdiv)}, 
   {\bff #1}, #2} 
\def\icrcparis#1#2{ 1981, in {\itt Proc. 17th ICRC(Paris)}, 
   {\bff #1}, #2} 
\def\icrcbang#1#2{ 1983, in {\itt Proc. 18th ICRC(Bangalore)}, 
   {\bff #1}, #2} 
\def\icrclajolla#1#2{ 1985, in {\itt Proc. 19th ICRC(La Jolla)}, 
   {\bff #1}, #2} 
\def\icrcmoscow#1#2{ 1987, in {\itt Proc. 20th ICRC(Moscow)}, 
   {\bff #1}, #2} 
\def\icrcadel#1#2{ 1990, in {\itt Proc. 21st ICRC(Adelaide)}, 
   {\bff #1}, #2} 
\def\icrcdub#1#2{ 1991, in {\itt Proc. 22nd ICRC(Dublin)}, 
  {\bff #1}, #2} 
\def\icrccalgary#1#2{ 1993, in {\itt Proc. 23rd ICRC(Calgary)}, 
  {\bff #1}, #2} 
\def\icrcrome#1#2{ 1995, in {\itt Proc. 24th ICRC(Rome)}, 
  {\bff #1}, #2} 
\def\icrcromepress{ 1995, {\itt Proc. 24th ICRC(Rome)}, in press}
\def\grl#1#2#3{ 19#1, {\itt G.R.L., } {\bff #2}, #3}
\def\jcp#1#2#3{ 19#1, {\itt J. Comput. Phys., } {\bff #2}, #3}
\def\JETP#1#2#3{ 19#1, {\itt JETP , } {\bff #2}, #3}
\def\JETPlet#1#2#3{ 19#1, {\itt JETP Lett., } {\bff #2}, #3}
\def\jgr#1#2#3{ 19#1, {\itt J.G.R., } {\bff #2}, #3}
\def\jpG#1#2#3{ 19#1, {\itt J.Phys.G: Nucl. Part. Phys., } 
     {\bff #2}, #3}
\def\mnras#1#2#3{ 19#1, {\itt M.N.R.A.S.,} {\bff #2}, #3}
\def\nature#1#2#3{ 19#1, {\itt Nature,} {\bff #2}, #3} 
\def\nucphysB#1#2#3{ 19#1, {\itt Nucl. Phys. B (Proc. Suppl.,} 
    {\bff #2}, #3} 
\def\pss#1#2#3{ 19#1, {\itt Planet. Sp. Sci.,} {\bff #2}, #3}
\def\pf#1#2#3{ 19#1, {\itt Phys. Fluids,} {\bff #2}, #3}
\def\phyrepts#1#2#3{ 19#1, {\itt Phys. Repts.,} {\bff #2}, #3}
\def\pr#1#2#3{ 19#1, {\itt Phys. Rev.,} {\bff #2}, #3}
\def\prD#1#2#3{ 19#1, {\itt Phys. Rev. D,} {\bff #2}, #3}
\def\prl#1#2#3{ 19#1, {\itt Phys. Rev. Letts,} {\bff #2}, #3}
\def\pasp#1#2#3{ 19#1, {\itt Pub. Astro. Soc. Pac.,} {\bff #2}, #3}
\def\revgeospphy#1#2#3{ 19#1, {\itt Rev. Geophys and Sp. Phys.,} 
   {\bff #2}, #3}
\def\rmp#1#2#3{ 19#1, {\itt Rev. Mod. Phys.,} {\bff #2}, #3}
\def\science#1#2#3{ 19#1, {\itt Science,} {\bff #2}, #3}
\def\sp#1#2#3{ 19#1, {\itt Solar Phys.,} {\bff #2}, #3}
\def\ssr#1#2#3{ 19#1, {\itt Space Sci. Rev.,} {\bff #2}, #3}
%

%rrr
\clearpage

\clearpage

%\centerline{\bf FIGURE CAPTIONS}\vskip15pt

\figcaption[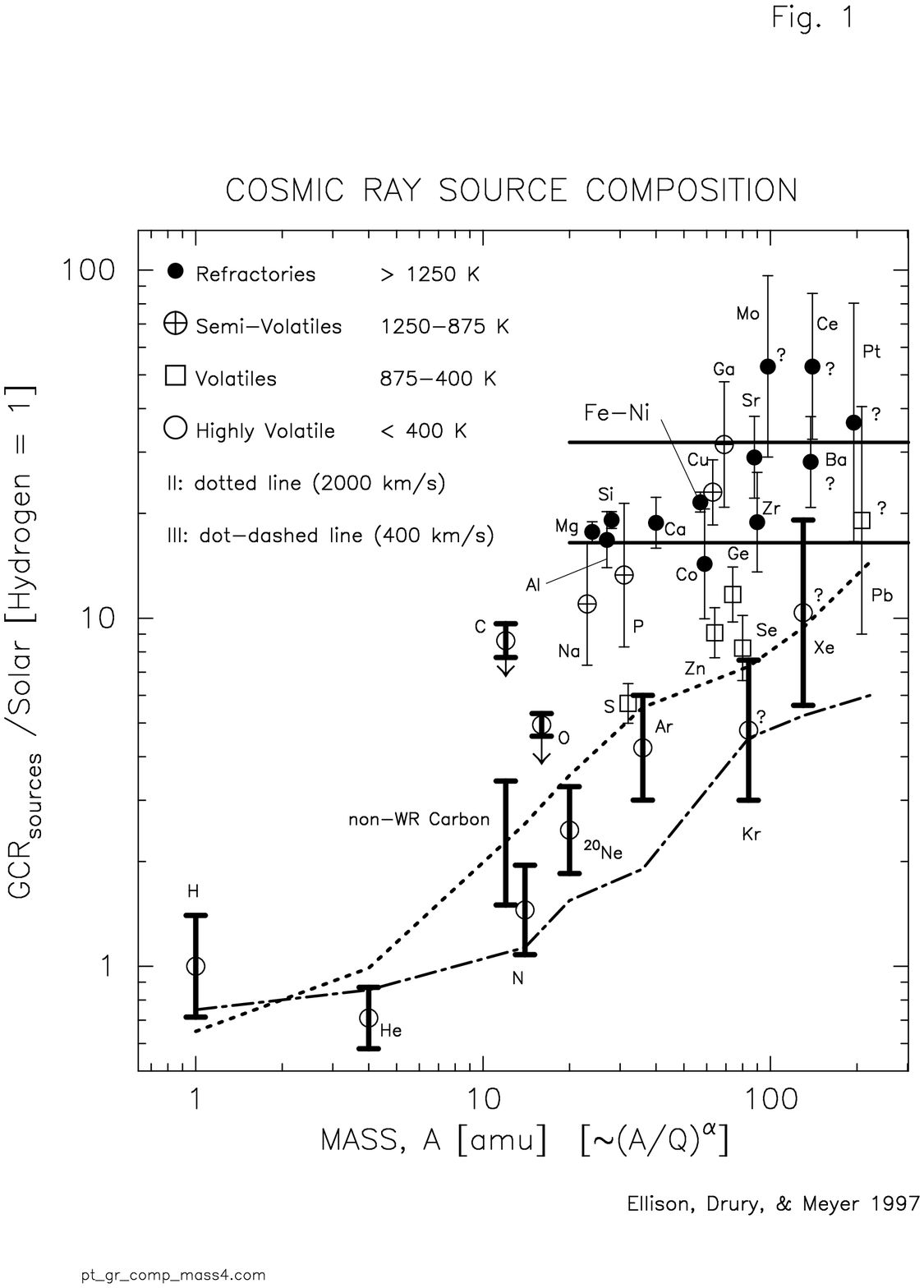]{Galactic cosmic ray source abundance relative to solar
abundance versus atomic mass number.  
% zzz
All values are measured relative
to cosmic ray hydrogen at a given energy per nucleon.  
% zzz
The elements are divided, on the basis of condensation temperature,
into refractories, semi-volatile, volatile, and highly volatile
groups. The refractories are essentially completely locked in grains
in the ISM, while the highly volatile elements are gaseous.  The
arrows on carbon and oxygen indicate that these elements have an
additional source from $^{22}$Ne-C-O-enriched Wolf-Rayet wind
material. Our estimate for the non-W-R contribution of carbon is
labeled.
% zzz
Our predictions
for the abundances of volatile elements from a high Mach number shock
model are shown as a dotted line, and for 
a lower Mach number model with a dot-dashed line.
% zzz
The horizontal solid lines
on the right side of the plot are limits on our predicted abundance of
iron and other refractory elements.  The label on the abscissa
[$\sim(A/Q)^\alpha$, where $\alpha$ is some unspecified constant] is a
reminder that, for most ionization models, $A/Q$ is a roughly
monotonically increasing function of the mass.  
% zzz 
We note that the abundances of Kr, Xe, Mo, Ba, Ce, Pt, and Pb relative
to Fe may contain systematic errors which are difficult to evaluate
(we indicate this with a `?' to the right of the point).
% zzz
For a complete
discussion of the observations, see Meyer, Drury, \& Ellison (1997)
(Paper I).                                         \label{fig:compmass}
}

\figcaption[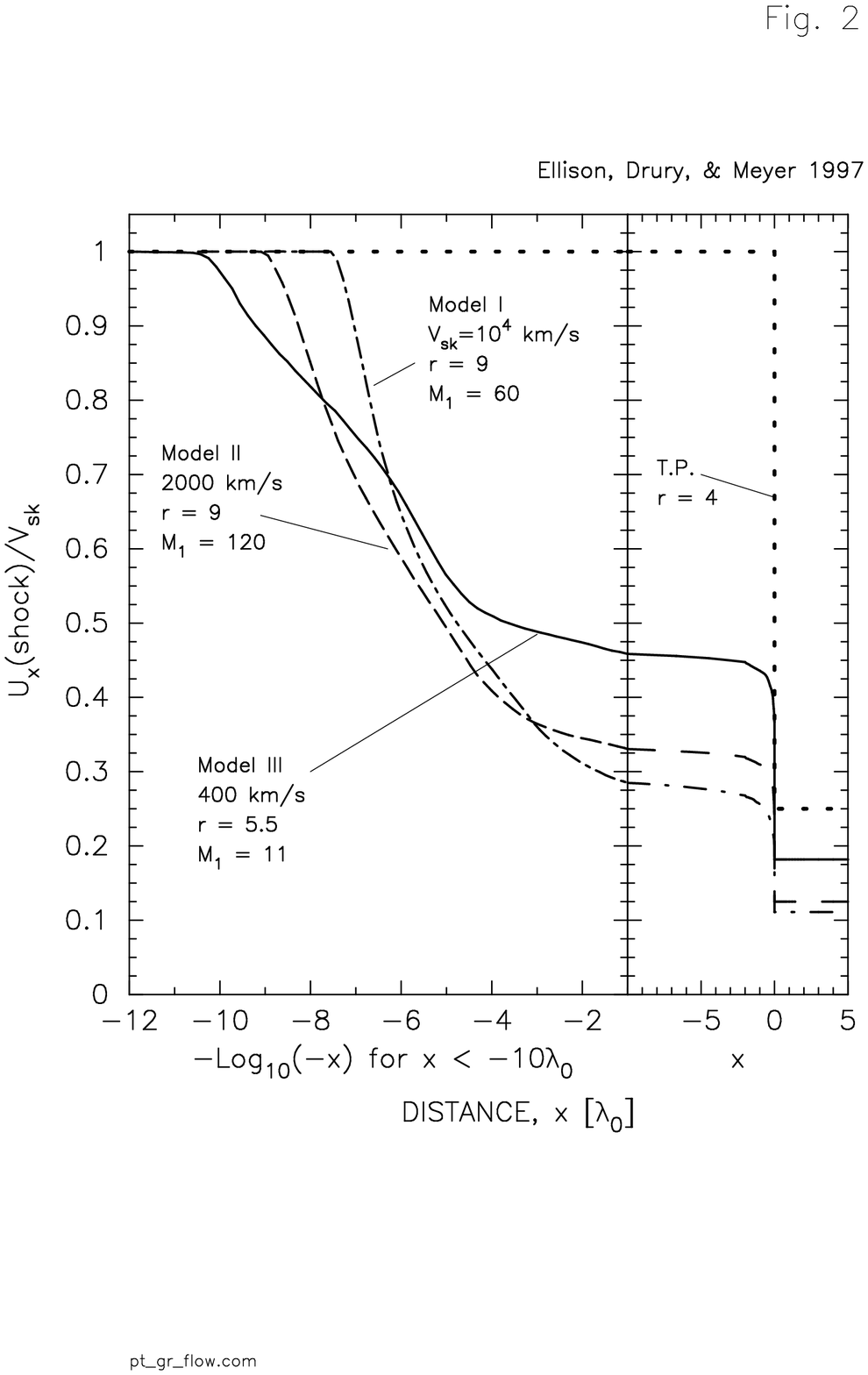]{
Average bulk flow speed in the shock frame, $\Ux$, versus distance
(\iec the shock structure) obtained from the \mc model. The distance
scale is logarithmic for $x<-10\Lz$ and linear for $x>-10\Lz$. The
vertical scale is in units of the far upstream speed, $\Vsk$, and
$\Lz= \eta \rgone$, where $\rgone$ is the gyroradius of a far upstream
proton with a speed equal to the shock speed, which varies for each
model. The compression ratios, $r$, in these nonlinear models depend
on the fraction of pressure carried by relativistic particles and on
the amount of energy escaping at the upstream free escape boundary
(FEB) and are always greater than the standard Rankine-Hugoniot value.
The dotted line shows a test-particle profile of a shock with $r= 4$.
Note that even though the nonlinear shocks are smoothed on the length
scales of the highest energy particles in the system, a well defined
subshock remains in all cases.  Our steady-state, plane-shock model is
such that the profiles are constant behind the shock (\ie $x>0$).
\label{fig:flow}
}

\figcaption[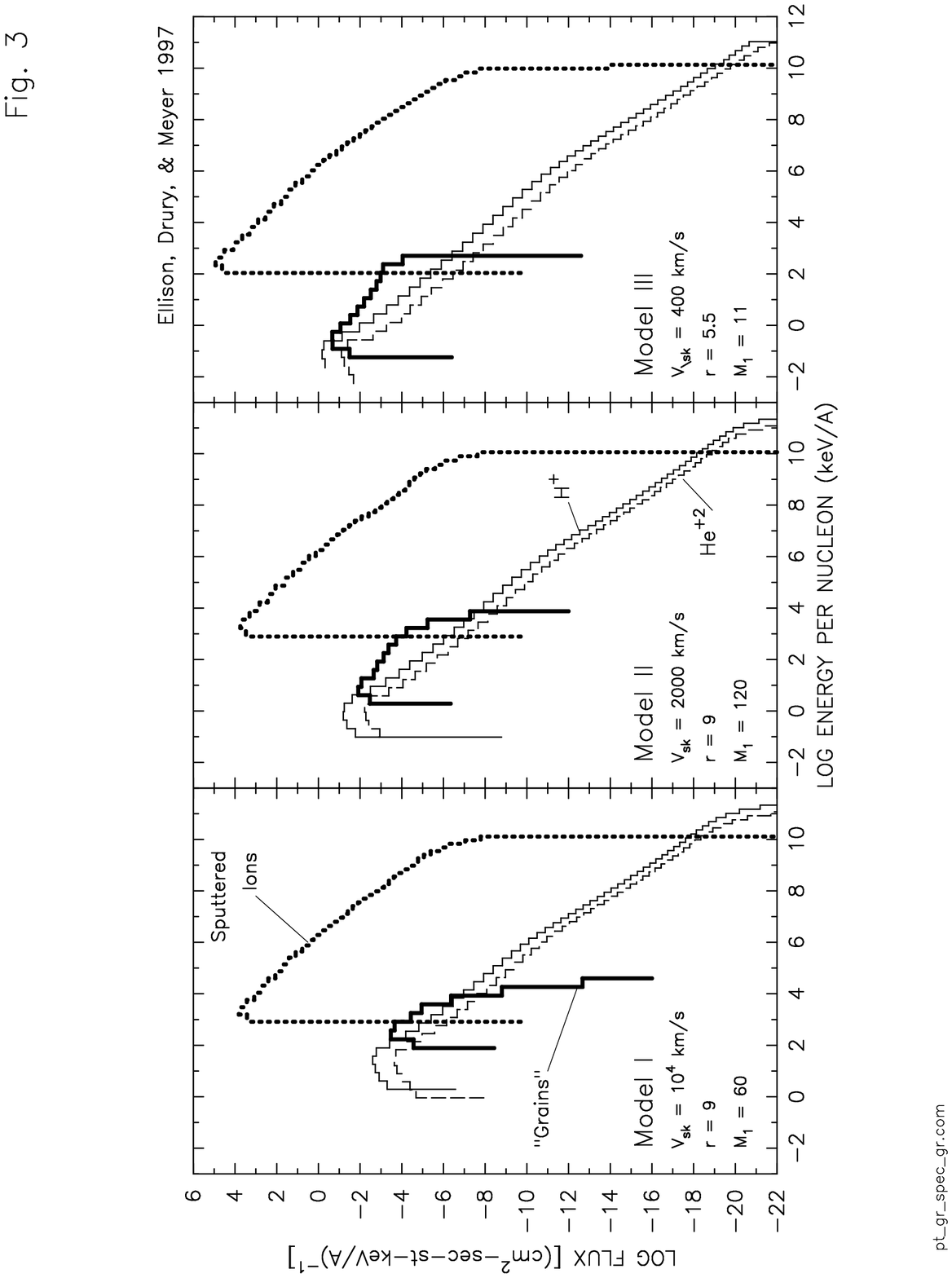]{
Differential flux spectra in energy per nucleon obtained from shocks
with $\Vsk= 10^{4}$ \kmps \ (Model I), $\Vsk= 2000$ \kmps \ (Model
II), and $\Vsk= 400$ \kmps \ (Model III).  The light solid curves are
the proton spectra, the light dashed curves are \Hetwo spectra, the
heavy solid curves are ``Fe grain'' spectra, and the heavy dotted
lines are Fe ions sputtered off the grains.  In all models, the far
upstream proton flux is normalized to one particle per cm$^2$ per sec,
$n_{\rm He}/n_{\rm H} = 0.1$ far upstream from the shock, and the
grains are test particles and are injected far upstream with the same
number density as protons.  At energies below the fall off produced by
frictional losses, grains experience a large enhancement over protons
at least in Models II and III.  All spectra here and elsewhere are
calculated in the shock frame at a position downstream from the shock.
%
% zzz - 7
%
Note that the grain quasi-thermal peak decreases in energy per nucleon
both absolutely and relative to the proton quasi-thermal peak as the
shock speed decreases. This reflects the fact that in all cases the
grains feel essentially the entire shock density jump and obtain an
energy per nucleon of $\sim \mp \Vsk/2$ after one shock crossing,
whereas the protons obtain a quasi-thermal peak determined by the subshock
strength.
%
% zzz - 7
%
\label{fig:diffspec}
}

\figcaption[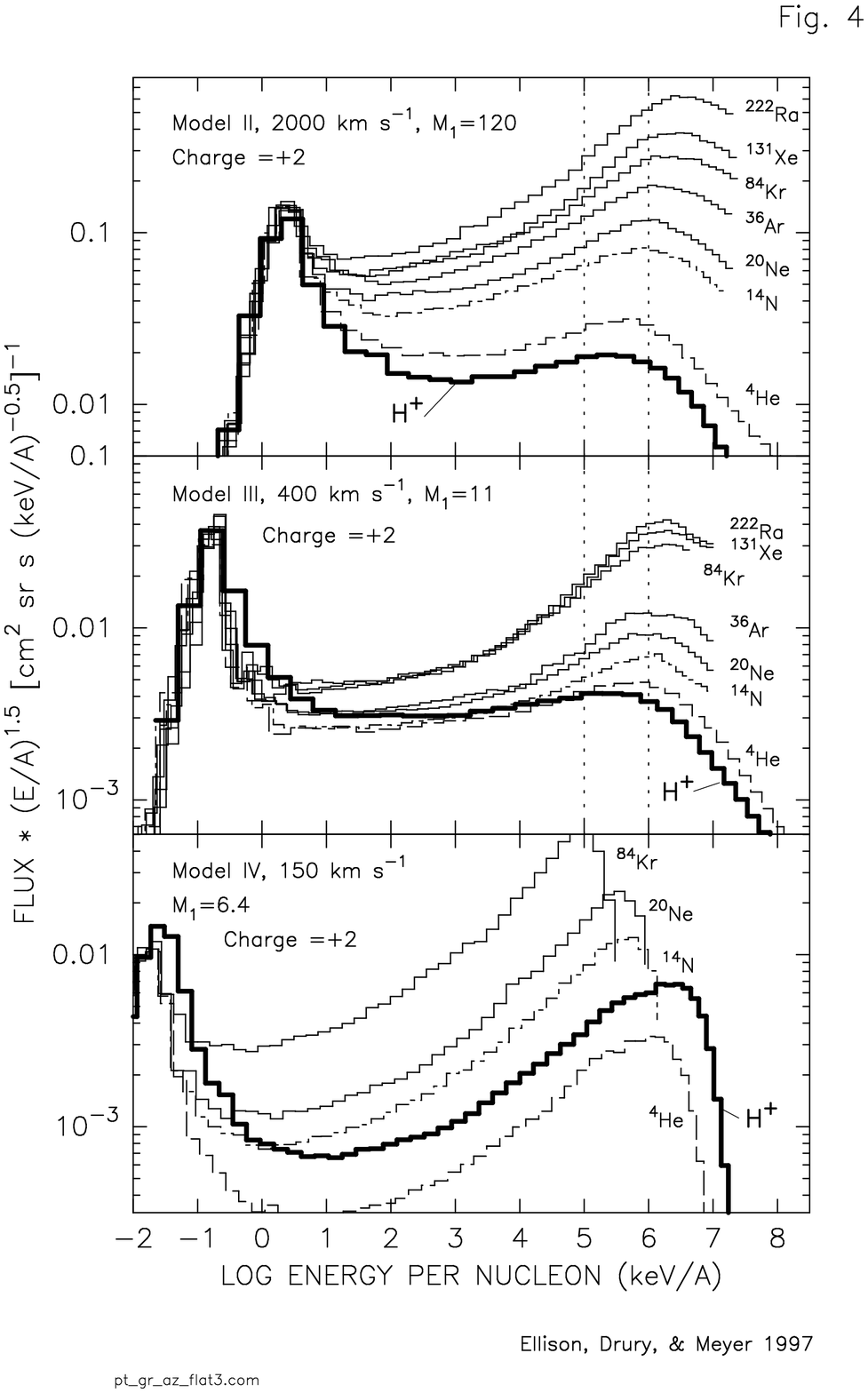]{
Differential flux spectra multiplied by $(E/A)^{1.5}$. All spectra are
normalized to one far upstream particle injected per cm$^2$ per second and
are obtained using smooth shock 
% zzz
profiles. The shock structures of Models II (top panel) and III
(middle panel) are shown in Figure~\protect\ref{fig:flow}. The vertical dotted
lines (Models II and III) show the energy per nucleon where the
abundance ratios shown in Figure~\protect\ref{fig:azratio} are calculated. For
computational reasons, Model IV has a lower maximum cutoff energy than
the other models. The low cutoff energy results in a high compression
ratio which causes the enhancements of heavy elements to be larger
than would be the case if a higher cutoff energy was used (see Ellison
\& Reynolds 1991 for details).
\label{fig:azflat}
}

\figcaption[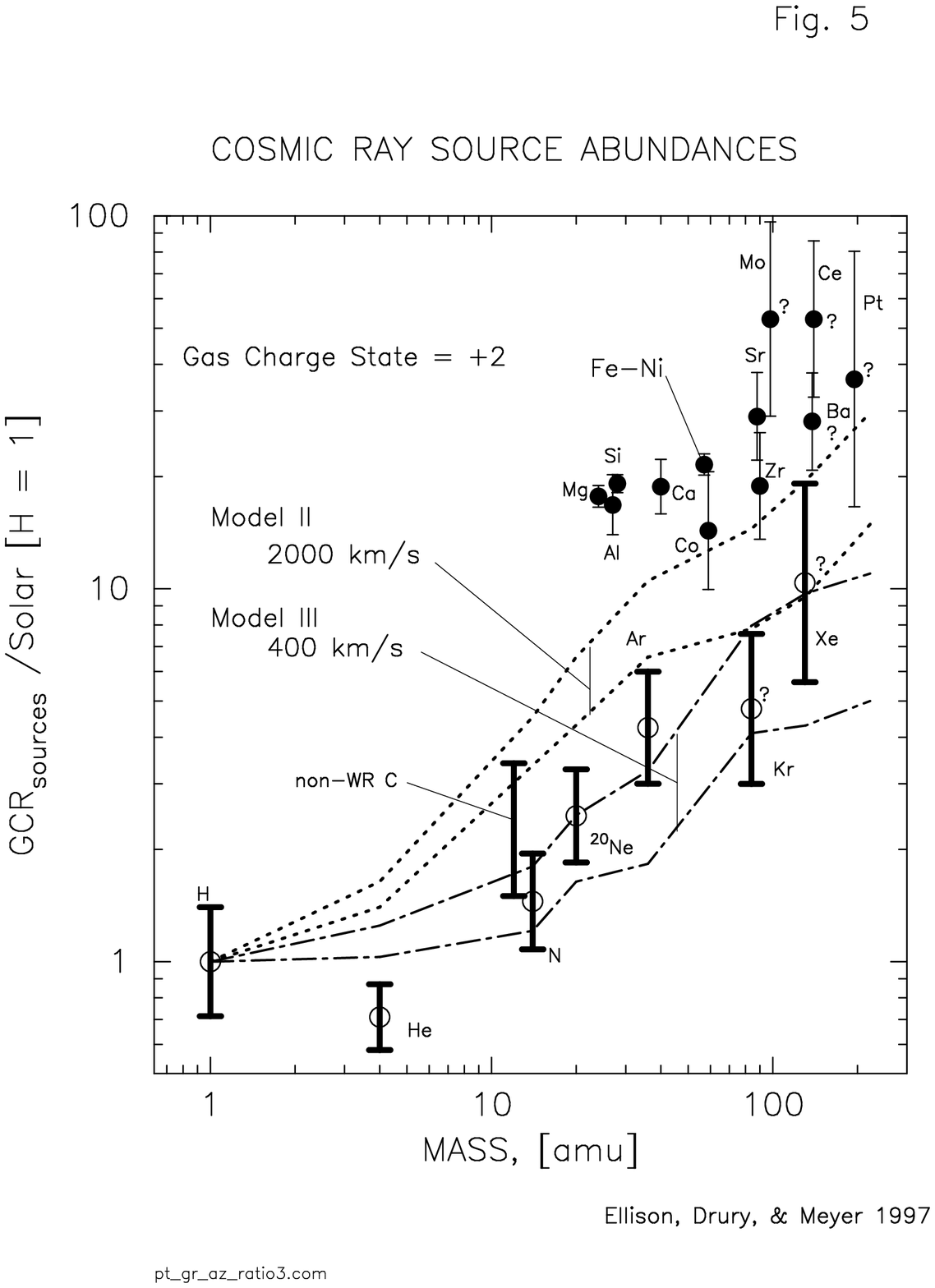]{
Volatile galactic cosmic ray source abundance relative to solar
abundance (open circles) versus atomic mass number.  As in
Figure~\protect\ref{fig:compmass}, the observed values are measured relative
to cosmic ray hydrogen at a given energy per nucleon.  Our estimate
for the non-W-R contribution of carbon is labeled.  The model
predictions, which assume a constant charge state of $+2$ for the
gaseous elements, are shown as dotted lines (Model II, $\Vsk=2000$
\kmps) and dot-dashed lines (Model III, $\Vsk=400$ \kmps). In each
case, the upper line was calculated at 1 GeV/A, the lower line was
calculated at 100 MeV/A, and the value for hydrogen was set to one.
These two energies bracket the range where we believe charge stripping
will produce similar $A/Q$'s for all heavier ions which precludes
further preferential acceleration.  For comparison, we also show the
refractory cosmic ray abundances (solid dots) to emphasize that
gas-phase elements and refractories can be cleanly separated in the
observations.
\label{fig:azratio}
}

\figcaption[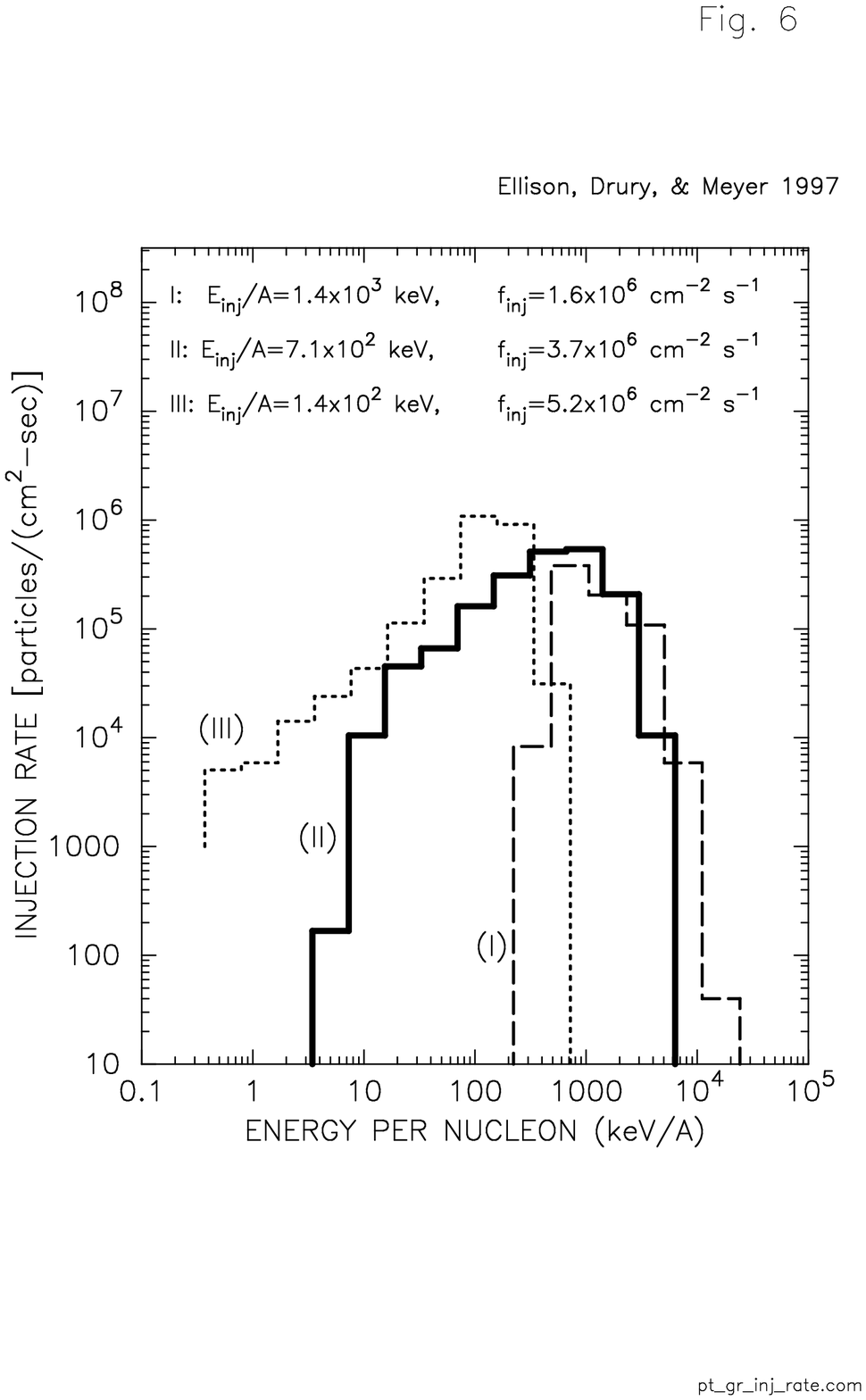]{
Injection rates for the superthermal sputtered Fe ions, $I(v)$, versus
energy obtained from the grain spectra shown in Figure~\protect\ref{fig:diffspec}. The
rates peak strongly just below the cutoff from direct losses allowing
a $\delta$-function approximation with the labeled values for the
injection of the sputtered products shown in Figure~\protect\ref{fig:diffspec}.
\label{fig:injrates}
}

\figcaption[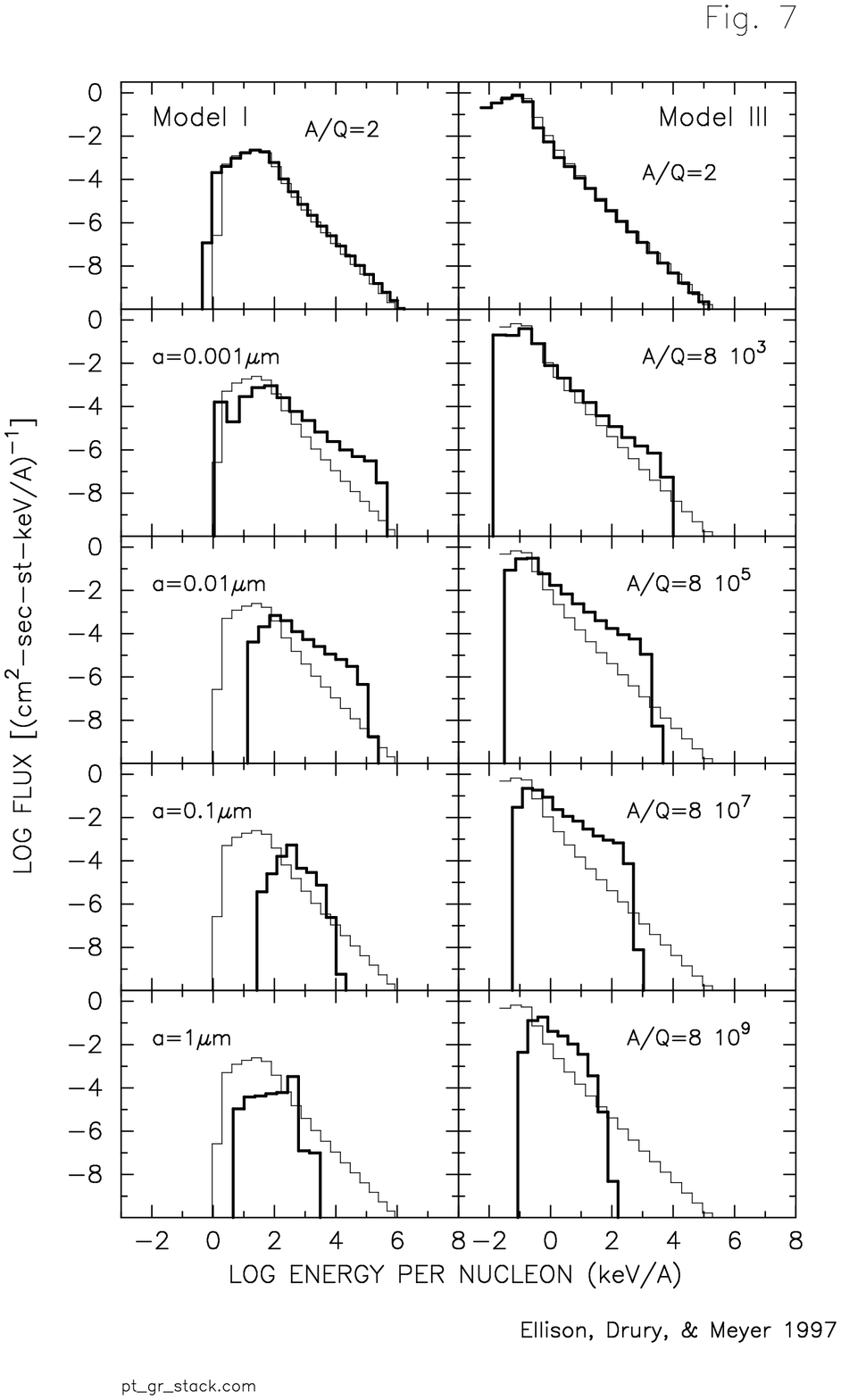]{
Differential flux spectra in energy per nucleon.  In all cases, the
light line is the low energy portion of the proton spectrum shown in
Fig.~\protect\ref{fig:diffspec}.  The heavy lines show grain spectra for various grain
sizes, $a$, for each horizontal pair of panels except for the top.
The $A/Q$ values are calculated for $\phi= 10$ V and $\mu= 56$.  Model
I has a shock speed, $\Vsk= 10^{4}$ \kmps, while Model III has $\Vsk=
400$ \kmps. The results for Model II are intermediate to these.
\label{fig:diffspecbig}
}

\figcaption[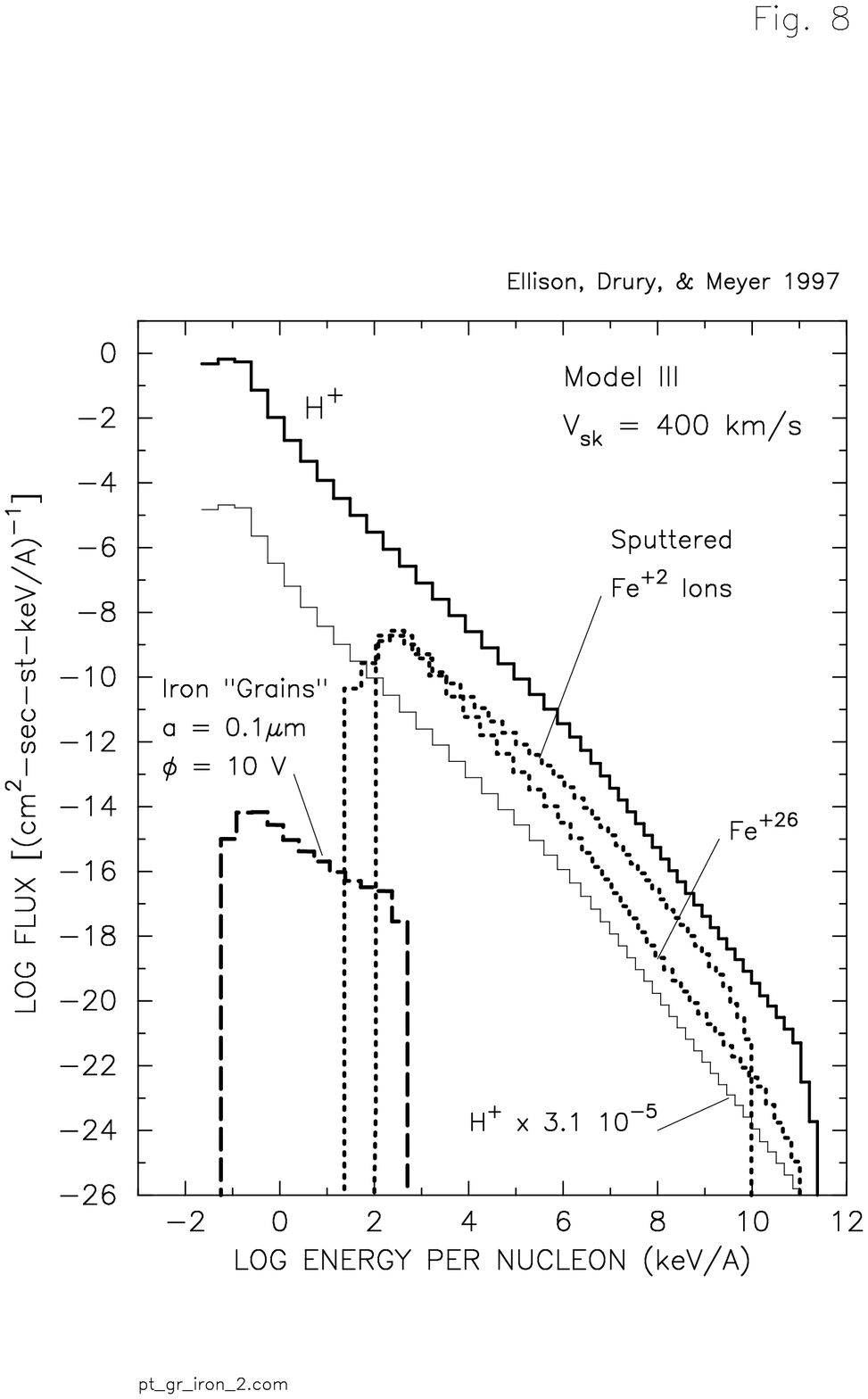]{
The heavy solid line shows the same proton spectrum as shown in the
right-hand panel of Figure~\protect\ref{fig:diffspec}. The light solid line is this
proton spectrum multiplied by $3.1\x{-5}$, i.e., scaled to the solar
Fe/H ratio. The dashed line shows the grain spectrum normalized to the
cosmic abundance of iron, i.e., the far upstream grain number density,
$\denGr=3.1\x{-5}\denH/10^9$ \pcc, where there are $10^9$
iron atoms per grain. The two dotted lines show the spectra of the
accelerated sputtered iron ions assuming constant charge states of
$+2$ and $+26$. Note that for either charge state, the sputtered iron
ends up with a flux greater than its cosmic abundance would suggest.
We consider the Fe$^{+26}$ flux the most realistic.
\label{fig:ironspec}
}

\figcaption[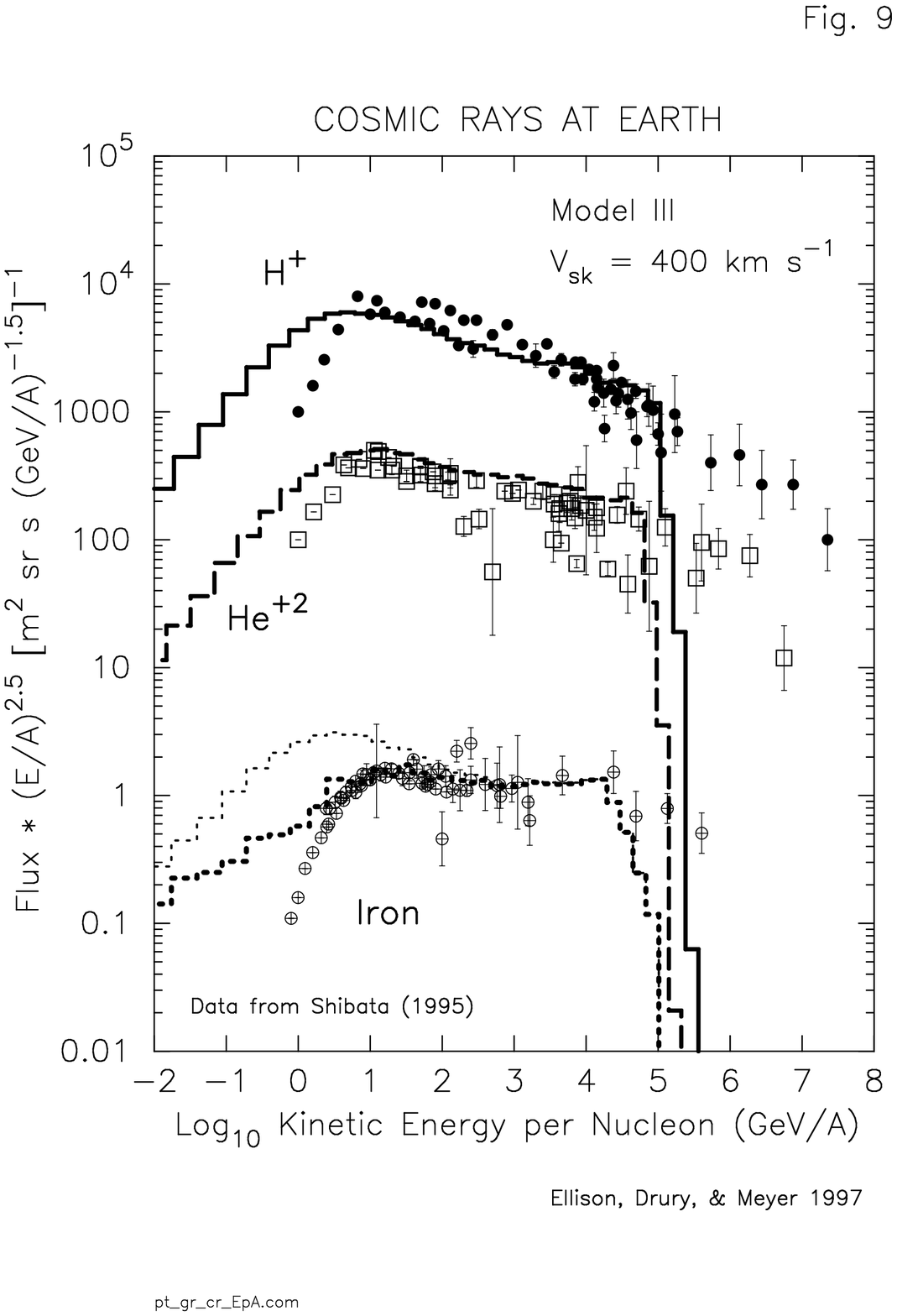]{
Cosmic ray spectra in energy per nucleon.  The cosmic ray data are
from the compilation of Shibata (1995) and have been multiplied by
$(E/A)^{2.5}$ to produce nearly flat spectra above $\sim 1$ GeV/A.
Note that in addition to being multiplied by $(E/A)^{2.5}$, the three
source model curves (protons, solid line; He$^{+2}$, dashed line;
Fe$^{+26}$, dotted lines) have been multiplied by $R^{-0.65}$ to mimic
rigidity dependent escape from the galaxy and corrected for nuclear
destruction during propagation.  In the model, both helium and iron
are injected at solar abundance, i.e., $\denHe / \denH = 0.1$ and
$\denFe / \denH = 3.1\x{-5}$. The normalization of the model proton
spectrum is varied to match the observations but the relative
normalization of helium to hydrogen and iron to hydrogen is fixed by
the model.
The light dotted line shows the iron
spectrum without correction for nuclear destruction.
The apparent differences in spectral shape between the uncorrected
iron and the other model
spectra, at fully relativistic energies, are mainly the result of poor
statistics.  
\label{fig:crdata}
}

\clearpage

\epsscale{0.75} % figures are a little large to fit on page!

\plotone{fig1.ps}

\clearpage
\plotone{fig2.ps}

\clearpage
\plotone{fig3.ps}

\clearpage
\plotone{fig4.ps}

\clearpage
\plotone{fig5.ps}

\clearpage
\plotone{fig6.ps}

\clearpage
\plotone{fig7.ps}

\clearpage
\plotone{fig8.ps}

\clearpage
\plotone{fig9.ps}

\end{document}